\documentclass[aps,prb,10pt,showpacs,superscriptaddress,twocolumn,amssymb,amsfonts]{revtex4-1}

\usepackage{amsmath}
\usepackage{bm}
\usepackage{graphics}
\usepackage{dsfont}
\usepackage{subfigure}		
\usepackage{epsfig}
\usepackage{hyphenat}
\usepackage{hyperref}		

\newcommand{\bra}[1]{ {\langle{#1}|} }
\newcommand{\ket}[1]{ {|{#1}\rangle} }
\newcommand{\braket}[2]{ {\langle{#1}|{#2}\rangle} }

\newcommand{\MxTwo}[2]{ {\begin{bmatrix} #1 \\ #2 \end{bmatrix}} }

\newcommand{\Tr}{ \operatorname{Tr} }
\newcommand{\Det}[1]{ \operatorname{Det}\left[{#1}\right] }
\newcommand{\sgn}{ \operatorname{sgn} }

\renewcommand{\Im}{ \operatorname{Im} }

\newcommand{\phd}{ {\vphantom{\dag}} }		
\newcommand{\vk}{ {\mathbf{k}} }
\newcommand{\vkpara}{ {\mathbf{k}_\parallel} }
\newcommand{\kperp}{ {k_\perp} }
\newcommand{\vp}{ {\mathbf{p}} }
\newcommand{\va}{ {\mathbf{a}} }
\renewcommand{\vr}{ {\mathbf{r}} }

\newcommand{\vb}{ {\mathbf{b}} }
\newcommand{\vh}{ {\mathbf{h}} }
\newcommand{\vGamma}{ {\bm\Gamma} }
\newcommand{\vbpara}{ {\vb^0_\parallel} }
\newcommand{\vbperp}{ {\vb^0_\perp} }
\newcommand{\vhpara}{ {\vh_\parallel} }
\newcommand{\vc}{ {\mathbf{c}} }
\newcommand{\vcperp}{ {\vc^0_\perp} }
\newcommand{\vbeta}{ {\bm\beta} }
\newcommand{\vvhat}{ {\hat{\mathbf{v}}} }

\newcommand{\ZZ}{ {\mathbb{Z}} }

\usepackage[usenames,dvipsnames]{color}			
\definecolor{purple}{rgb}{0.5,0,0.5}

\begin{document}
\title{Edge states and the bulk-boundary correspondence in Dirac Hamiltonians}

\author{Roger S.~K.~Mong}
	\affiliation{Department of Physics, University of California, Berkeley, CA 94720, USA}
\author{Vasudha Shivamoggi}
	\affiliation{Department of Physics, University of California, Berkeley, CA 94720, USA}
\date{\today}

\begin{abstract}


We present an analytic prescription for computing the edge dispersion $E(k)$ of a tight-binding Dirac Hamiltonian terminated at an abrupt crystalline edge.
Specifically, we consider translationally invariant Dirac Hamiltonians with nearest-layer interaction.
We present and prove a geometric formula that relates the existence of surface states as well as their energy dispersion to properties of the bulk Hamiltonian.
We further prove the bulk-boundary correspondence between the Chern number and the chiral edge modes for quantum Hall systems within the class of Hamiltonians studied in the paper.
Our results can be extended to the case of continuum theories which are quadratic in the momentum, as well as other symmetry classes.

\end{abstract}


\maketitle

\section{Introduction}

Topological order is responsible for interesting new states of matter that do not fit into the standard symmetry-breaking picture.~\cite{XGWenReview95}
For decades, the Landau paradigm successfully described systems by looking at the underlying symmetries, with phase transitions occurring between phases with different symmetries.
However the integer quantum Hall (IQH) effect showed this approach to be incomplete, since it exhibits transitions between phases of the same symmetry.  These phases are instead distinguished by topological order, with gapless modes localized at domain walls between regions of different topological order.

For systems with a non-zero bulk gap at all points in the Brillouin zone, it is possible to define a topological invariant of the Hamiltonian.
Systems with non-trivial topological invariants are termed topological insulators (TI) and topological superconductors.%
~\cite{*[{For a review of topological insulators, the reader may consult }][] MooreTIBirth, *HasanKaneReview, *QiSCZhangTITSCReview, SFRLClassification3D, KitaevClassification}
The invariants are robust to smooth deformations that do not close the bulk gap and underlie the precise quantization of response functions in topologically ordered systems.
This was first realized in IQH states, where the Hall conductivity can be expressed as the first Chern number of the $U(1)$ vector bundle of Bloch states.~\cite{TKNN,ASSHomotopy,SimonBerry83}
For time-reversal invariant (TRI) systems in two-dimensions (2D), a $\ZZ_2$ topological invariant distinguishes between the vacuum (trivial phase) and the quantum spin Hall (QSH) state.~\cite{KaneMeleQSH,KaneMeleZ2,BernevigSCZhangHgTe,MolenkampHgTe}
In three-dimensions (3D), there are four $\ZZ_2$ invariants describing TRI systems, one of which distinguishes between the vacuum and a strong topological insulator and is robust to disorder.~\cite{FuKaneMeleTI3D,MooreBalents06,RoyQSH3D,HsiehTIDirac,HasanBi2Se3,YLChenBi2Te3}
The quantized magnetoelectric response may be written in terms of this $\ZZ_2$ invariant.~\cite{QiHughesZhangTFT,EssinMPAxion09}

Although a topological invariant is an abstract quantity defined for a fully periodic system, it is manifested physically as mid-gap surface states.~\cite{SFRLClassificationSurface}
In IQH systems, the quantized Hall conductance can be formulated in terms of the the number of chiral edge states.~\cite{LaughlinQHE81,HalperinQHE82}
Similarly, the 2D/3D $\ZZ_2$ invariant for TRI systems determines whether there are an odd or even number of helical modes/Dirac cones at a given edge or surface.~\cite{FuKaneTRPZ2}
In the cases above, the edge states smoothly connect the bulk valence and conduction bands and the number of such modes is protected by the topological invariant: they cannot be deformed into a single bulk band unless the bulk gap closes.  By contrast, edge modes in an ordinary system do not traverse the bulk gap and are thus susceptible to localization by disorder.
For superconducting systems, the topological invariants determine the number of Majorana modes localized at the edge or in vortices.~\cite{ReadGreenp+ipFQHE,KitaevWireMajorana01}
These states are at zero energy and are protected by particle-hole symmetry and the superconducting gap.  Systems such as $p+ip$ superconductors (SC) in the ``weak pairing phase'' or SC-TI heterojunctions can support Majorana modes which obey non-Abelian statistics.~\cite{FuKaneSCProximity08,MooreReadNonabelion91,IvanovNonAbelian01}

The goal of this paper is to derive a rigorous connection between the bulk invariants and the surface dispersion.
A heuristic way to understand this bulk-boundary correspondence is as follows.
Consider a domain wall between two bulk insulators with suitably defined topological invariants that take the values $\nu _L$ and $\nu _R \neq \nu _L$ in some regions.  Since the value of the invariant cannot change for finite energy gap, this means the bulk gap closes at some interface.  Mid-gap excitations can thus exist, but they are confined to the interface by the bulk gap in the other regions.
This argument applies to domain walls between regions with different values of the invariant, of which an edge is a special case where one of the regions is the vacuum (trivial phase).~\cite{JackiwRebbi76,TeoKaneTopologicalDefects} 

In light of recent interest in topological insulators and superconductors, it would be useful to formalize the relation between bulk topological quantities and properties of mid-gap edge states.
This connection has been proven specifically for IQH states on a square lattice by deriving a winding number for the edge states.~\cite{HatsugaiIQHEprl93}
Another approach using twisted boundary conditions has the advantage of including interactions and disorder, but cannot prove that the states exist at an open boundary.~\cite{XLQiWuSCZhangBulkEdge06}

There has also been recent progress on analytic solutions of edge states in topological insulators~\cite{MaoAnalyticEdgeModeTI10,SCZhangBi2X3DiracCone,SQShenZ2DiracEquation10} and topological superconductors.~\cite{SchnyderAndreevSurfaceTSC10}
These calculations are often based on models using a specific Dirac Hamiltonian.
Dirac systems are ubiquitous in condensed matter and particle physics systems and give rise to many exotic states.  
For example, every single-particle topologically ordered system can be realized with a Dirac Hamiltonian.%
~\cite{KitaevClassification,RyuLudwigDimHeir10}
They are used to model a variety of systems including polyacetylene, quantum Hall insulators, graphene, topological insulators and superconductors, \textit{etc.}%
~\cite{SchriefferPolyacetylene79,HaldaneQHE88,KaplanChiralFLattice92,FujitaGrapheneEdge96,ReadGreenp+ipFQHE,KitaevWireMajorana01,FuKaneMeleTI3D,CXLiuModelHamTI10}

In this paper we deepen the understanding of surface states by deriving their dispersion, effective theory, and chiral properties.
Our work applies specifically to tight-binding Dirac Hamiltonians with nearest-layer interaction.
For these systems we present a prescription for the edge states spectrum and prove the bulk-boundary correspondence.
In addition, we derive a simple geometric method to calculate the energies and penetration depth of the edge states analytically.

The organization of the paper is as follows.
In section~\ref{sec:NNBulkEdge}, we introduce the bulk quantities of a lattice Hamiltonian that determine topological behavior.
In section~\ref{sec:LatticeProof}, we state and prove the two main results of the paper, Theorem~1 relating the parameters of the bulk Hamiltonian to the surface spectrum in a geometric way, and Theorem~2 proving the bulk-boundary correspondence between chiral edge states and the Chern number.
In section~\ref{sec:LatticeExamples}, we demonstrate the range of applicability of our theorems and give examples of topologically ordered systems.
We also show how the bulk $\ZZ_2$ invariant for a time\hyp{}reversal symmetric insulator relates to the number of surface Dirac cones in example~\ref{sec:LatticeExampleCubicTI}.
In section~\ref{sec:ContinuumBulkEdge}, we extend the results from lattice Hamiltonians to continuum quadratic Hamiltonians, with discussions on its implications.
In closing, we discuss the possible extension of the work to other classes of topological superconductors beyond IQH and TRI systems.

\section{Characterization of the nearest-layer Hamiltonian}
\label{sec:NNBulkEdge}

To study a system with edges, consider a 2D/3D crystal that terminates on a line/surface.  Translational symmetry is thus broken in the direction normal to the edges.  However, we assume it is unbroken parallel to the surface, and the corresponding momentum $\vkpara$ is a good quantum number.
In this way, any higher dimensional system can be decoupled into a family of one-dimension (1D) problems parameterized by $\vkpara$.

The Dirac Hamiltonian in momentum space $H(\vk)$ can always be expressed as a linear combination of gamma matrices, $H(\vk) = \vh(\vk) \cdot \vGamma$.
Here $\vGamma$ is a vector of the hermitian gamma matrices (independent of $\vk$)
which satisfy the Clifford algebra $\Gamma^i\Gamma^j + \Gamma^j\Gamma^i = 2\delta^{ij}$.
$\vh$ is a real vector that maps the Brillouin zone to a closed curve in a $g$-component vector space, where there are $g$ gamma matrices $\Gamma^i$.
The Pauli matrices are examples of gamma matrices: any $2\times2$ traceless matrix can written as a $\vh\cdot\bm\sigma$ where $\vh$ is a 3-component vector and $\bm\sigma = (\sigma^x,\sigma^y,\sigma^z)$.

Squaring the Hamiltonian gives $H^2 = \left( \vh\cdot\vGamma \right)^2 = |\vh|^2$.  The eigenvalues of $H$, given by $E(\vk) = \pm |\vh(\vk)|$, can be thought of as the distance of the vector $\vh$ to the origin.  If $H(\vk)$ describes a band insulator with a bulk gap, then the locus of points traced by $\vh(\vk)$ never intersects the origin.

Let us consider Dirac Hamiltonians with coupling between neighboring layers:
\begin{align}
	\mathcal{H} & = \sum_{n, \vkpara} \Psi^\dag_{n,\vkpara} \vGamma \cdot
		\left[ \vb\,\Psi_{n-1, \vkpara} + \vb^0\,\Psi_{n, \vkpara} + \vb^\ast\,\Psi_{n+1, \vkpara}\right] ,
	\label{eq:LatticeHopping}
\end{align}
where $n$ labels the layers.
Both $\vb$ and $\vb^0$ are dependent on $\vkpara$ but we will not write this dependence explicitly.
$\Psi_{n, \vkpara}$ is a vector of quasiparticle annihilation operators at layer $n$ that captures all the degrees of freedom (\textit{i.e.} spin, pseudospin) at every site.
Fourier transforming ($n \rightarrow \kperp$) in the direction away from the edge, the bulk Hamiltonian becomes 
\begin{align}
	\mathcal{H} & = \sum_{\kperp, \vkpara}
 			\Psi^\dag_\vk \left[ \vh(\vk)\cdot\vGamma \right] \Psi_\vk ,
	\label{eq:LatticeMomentumHam}
\end{align}
with
\begin{align}
	\vh(\vk)
		& = \vb e^{-i\kperp} + \vb^0 + \vb^\ast e^{i\kperp}			\notag
		\\	& = \vb^0 + 2\vb^r \cos \kperp + 2\vb^i \sin \kperp,
	\label{eq:LatticehLoop}
\end{align}
where $\vb^r$ and $\vb^i$ are the real and imaginary components of the vector $\vb$ respectively.
We point out that $\vb$ and $\vb^0$ are independent of $\kperp$

The curve traced out by $\vh(\vk)$ for fixed $\vkpara$ is an ellipse living in the plane spanned by $\vb^r$ and $\vb^i$.  $\vb^0$ can be decomposed into a component $\vbperp$ that is normal to this plane, and $\vbpara$ that lies within it.  $\vbpara$ shifts the ellipse within the plane, while $\vbperp$ lifts it uniformly.
It will be useful to define
\begin{align}
	\vhpara(\vk) = \vbpara + 2\vb^r \cos\kperp + 2\vb^i \sin\kperp
	\label{eq:hProjection}
\end{align} 
as the projection of $\vh(\vk)$ on to the 2D plane spanned by $\vb^r$ and $\vb^i$.  Note that this plane contains the origin, while the plane containing $\vh$ is offset from the origin by the vector $\vbperp$.  Since $\vhpara$ maps the Brillouin zone to closed curves, it can be divided into two classes: ellipses that enclose the origin, and ellipses that do not.

\section{Edge state and bulk-boundary theorems}
\label{sec:LatticeProof}

As we shall prove in this section, the behavior of $\vh(\kperp)$ completely determines the topological nature of the system and holds the key to understanding the relation between existence of edge states and bulk topological invariants.
This section contains the main result of the paper, where we prove two theorems, one relating the spectrum of edge states to $\vh$, the other connecting $\vh$ to a bulk topological invariant.

\subsection{Edge state energy}
\label{sec:hEdgeStates}

\textbf{Theorem 1a}.
\textit{%
The system has mid-gap edge states if and only if $\vhpara(\kperp)$ encloses the origin.%
}

\textbf{Theorem 1b}.
\textit{%
The energies of the edge states are given by the distance from the origin to the plane containing $\vh$, \textit{i.e. $E_s = \pm|\vbperp|$}.
When the gamma matrices are the Pauli matrices, the energy of the left edge state (semi-infinite slab with $n > 0$) is given by:
$E_s = \vb^0 \cdot \frac{ \vb^r\times\vb^i }{ |\vb^r\times\vb^i| }$.%
}

\begin{figure}[tb]
	\includegraphics[width=0.32\textwidth]{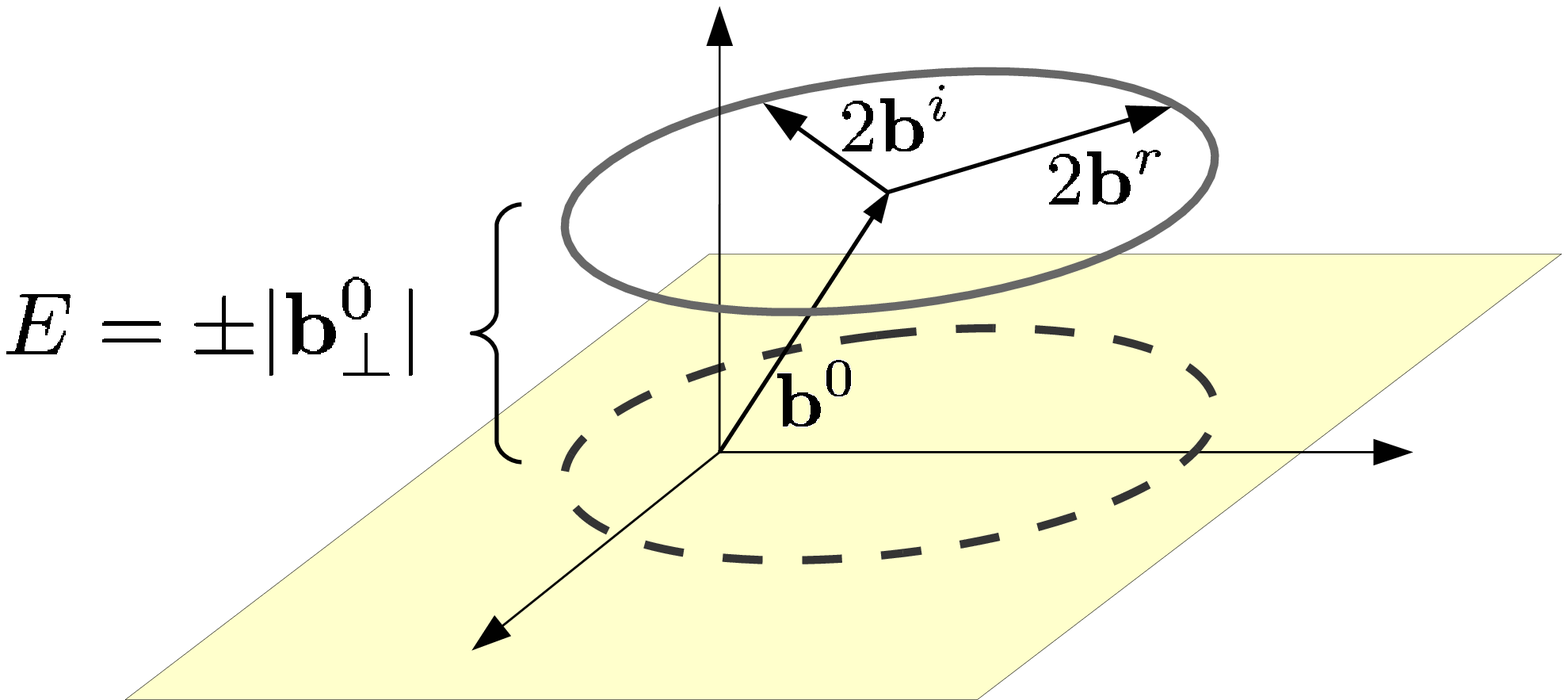}
	\caption{%
		An illustration of Theorem~1.
		The gray ellipse is traced out by $\vh(\kperp) = \vb^0 + 2\vb^r\cos\kperp + 2\vb^i\sin\kperp$ for a fixed parallel momentum $\vkpara$ [Eq.~\eqref{eq:LatticehLoop}].
		The dotted ellipse ($\vhpara$) is $\vh$ projected on to the plane spanned by $\vb^r$ and $\vb^i$.
		The displacement of the ellipse $\vh$ from the dotted ellipse $\vhpara$ is given by $\vbperp$, the component of $\vb^0$ perpendicular to this plane.
		Theorem 1 says that an edge state exists if and only if the dotted ellipse encloses the origin (which holds true for the diagram above), and its energy is determined by the displacement $|\vbperp|$.
	}
\end{figure}


Here we only sketch the main ideas behind two equivalent proofs of the theorem, leaving the full details to the Appendices.
We present two approaches to this problem: a proof utilizing Green's functions~\cite{ButtnerShockleyStates72} (\textit{c.f.} App.~\ref{sec:ProofGF}) and a proof via transfer matrices~\cite{DHLeeSurfaceBandTM81} (\textit{c.f.} App.~\ref{sec:ProofTM}).
In this section, we consider one block of the decoupled system corresponding to fixed $\vkpara$.

We begin by writing the Green's function of the system without edges, where the full translational invariance makes a momentum space representation possible.
A system with edges is then created from the fully periodic system by deleting the couplings between one pair of neighboring sites.  
The poles in the Green's function $G(E)$ at mid-gap energies $E$ indicate the presence of edge states.

The bulk Green's function is given by
\begin{align}
	G_0 \left( E; \kperp \right)
		= \sum_i {\frac{ \ket{\psi_i} \bra{\psi_i} }{E - E_i}}
	,	\label{eq:GFdefgreenmat}
\end{align}
where $i$ sums over the energy eigenstates of $H(\kperp)$.
Since we are interested in a boundary localized in real space, it is necessary to Fourier transform the bulk Green's function.  For a system of size $L$ this results in a $L \times L$ block matrix $G_0(E;y_i,y_j)$, where each block corresponds to mixing between the lattice sites $y_i$ and $y_j$.

Next we write an expression $V$ for the terms in the Hamiltonian that create the boundary by subtracting the hopping terms between sites $y_\textrm{left}$ and $y_\textrm{right}$.
For models with nearest neighbor interactions, the only non-zero matrix elements of $V$ are those between $y_\textrm{left}$ and $y_\textrm{right}$.
The Dyson equation gives an exact expression for the open boundary Green's function $G$ in terms of the bulk Green's function $G_0$ and the cuts $V$ needed to take the system from one geometry to the other:
\begin{align}
	G(E) = \big( I - G_0(E)V \big)^{-1} G_0(E) .
	\label{eq:GFdyson}
\end{align}
The poles of $G(E)$ occur when the edge state energy satisfies $\Det{I - G_0(E)V} = 0$.
If an edge state exists with wavefunction $u$, it must satisfy $(I - G_0V) u = 0$.
We note that this is the same as the Schr\"odinger equation $(E - H_0 - V)u = 0$.  The benefit of the Green's function formalism is that it reduces the problem to only edge degrees of freedom and enables an analytic solution.
This implies the following two statements:
\begin{subequations}
\begin{align}
	\int\! d\kperp \frac{\vhpara}{|\vhpara|^2} & = 0 ,
	\label{eq:GFintgeo1}\\
	\int\! d\kperp \, e^{i\kperp} \frac{\vhpara \cdot \vb^\ast}{|\vhpara|^2} & = \pi .
	\label{eq:GFintgeo2}
\end{align}  
\end{subequations}
These conditions are satisfied if and only if $\vhpara$ encloses the origin, and the edge mode energy is given by $\pm |\vbperp|$, where the sign is given by the orientation of $\vhpara$.


To prove Thm.~1 using transfer matrices, we consider a semi-infinite system with unit cells labeled by $n = 1,2,3,$ \textit{etc}.
We seek a solution $\psi_n$ to the single-particle Schr\"odinger equation:
\begin{align}
		\vb\cdot\vGamma \psi_{n-1} + \vb^0\cdot\vGamma \psi_n
				+ \vb^\ast\cdot\vGamma \psi_{n+1} & = E \psi_n
			\label{eq:TMRecurs}
\end{align}
for $n > 1$.
At the edge site $n = 1$, we have $\vb^0\cdot\vGamma \psi_1 + \vb^\ast\cdot\vGamma \psi_2 = E \psi_1$.  This condition is enforced by applying Eq.~\eqref{eq:TMRecurs} for $n=1$ but stipulating that $\psi_0 = 0$.
The recursion relation \eqref{eq:TMRecurs} relates $\psi_{n+1}$ to $\psi_n$ and $\psi_{n-1}$.  
Hence given $\psi_1$ (and $\psi_0 = 0$), we can recursively calculate all of $\psi_n$ and construct the wavefunction.

An edge state requires $\psi_n$ to be exponentially decaying as $n$ increases, hence the solution $\psi$ takes the form:
\begin{align}
	\psi_n = u_a \lambda_a^n + u_b \lambda_b^n ,
		\label{eq:TMWavefunction}
\end{align}
where $u_a = -u_b$, and $\lambda_a, \lambda_b$ are complex with $|\lambda_a|, |\lambda_b| < 1$.
Algebraically, this is equivalent to having $E = \pm|\vbperp|$ and finding two roots within the unit circle of the functions $L(\lambda)$ or $\bar{L}(\lambda)$, defined as
\begin{align}\begin{split}
		L(\lambda) & = \vhpara( -i \ln \lambda ) \cdot (\vvhat_1 + i\vvhat_2) ,
	\\	\bar{L}(\lambda) & = \vhpara( -i \ln \lambda ) \cdot (\vvhat_1 - i\vvhat_2) ,
\end{split}\end{align}
where $\vvhat_1,\vvhat_2$ are two orthonormal vectors that live in the plane of $\vhpara$.  
When $\lambda = e^{i\kperp}$ lies on the unit circle, $L(\lambda)$ and $\bar{L}(\lambda)$ trace out the ellipse $\vh(\kperp)$ in the complex plane clockwise and counterclockwise, respectively.
Because of this property, the number of times $\vh(\kperp)$ wraps the origin determines the number of zeroes of $L(\lambda)$ and whether the two solutions $\lambda_{a,b}$ in \eqref{eq:TMWavefunction} exists.
In Appendix~\ref{sec:ProofTM}, we provide the full details bridging these steps, and also compute the sign of the edge state energy as well as their penetration depth.

\subsection{Bulk Chern number and chiral edge correspondence}
\label{sec:BulkChiralChern}

In this section we prove Theorem~2, relating the bulk Chern number $\nu$ with the number of chiral edge modes for $2\times2$ Hamiltonians.

\textbf{Theorem 2}.
\textit{%
A chiral edge mode exists for a 2D bulk insulator if the bulk has a non-zero Chern number, \textit{i.e.} $\vh(\vk)$ wraps the origin.
The number of chiral edge modes, counterclockwise minus clockwise, is given precisely by the Chern number.%
}

When the irreducible representation of $\vGamma$ are $4\times4$ or larger, it can be shown that the Chern number is always zero.  The edge states of any surface always appear in pairs with energy $+E_s$ and $-E_s$ and so the number of clockwise and counterclockwise chiral modes are always equal.
We are particularly interested in $2\times2$ Hamiltonians because they can have nonzero Chern numbers and support chiral modes.

Consider an insulator in two dimensions whose Hamiltonian is written as a $2\times2$ traceless matrix: $H(k_x,k_y) = \vh(k_x,k_y) \cdot \bm\sigma$.
Because the bulk gap of an insulator is non-zero, $\vh$ is non-zero at all points in the Brillouin zone.
Hence $H(\vk)$ is a map from the Brillouin zone (torus) to a set of non-zero vectors with 3 components ($\mathbb{R}^3-\{0\}$), and such maps can be characterized by a $\nu\in\ZZ$ topological invariant, known as the
Chern number.~%
	\footnote{
	Technically the Chern number is not defined for the map $\vh:T^2 \rightarrow \mathbb{R}^3-\{0\}$.
	However, we can compose $\vh$ with the deformation retract $r: \mathbb{R}^3-\{0\} \rightarrow S^2 = \mathbb{C}\mathrm{P}^1$ and the inclusion map $i:\mathbb{C}\mathrm{P}^1 \rightarrow \mathbb{C}\mathrm{P}^\infty$ to make the Chern number (first Chern class) well defined: $\varphi = i \circ r \circ \vh : T^2 \rightarrow \mathbb{C}\mathrm{P}^\infty$.
	What it boils down to is that we are calling the induced map between the cohomology classes $\vh^\ast: H^2(\mathbb{R}^3-\{0\}) \rightarrow H^2(T^2)$ the Chern number.%
	}
Hamiltonians with different Chern numbers $\nu$ cannot be deformed into one another without closing the bulk gap.
In this context, the invariant $\nu$ determines the number of times the torus $\vh(\vk)$ wraps around the origin.

To examine the edge states for an arbitrary edge, say one parallel to $\hat{y}$, we analyze the spectrum as a function of $k_\parallel=k_y$.
The torus $\vh(\vk)$ can be divided into a family of loops $\vh(k_x)|_{k_y}$, each at a fixed value of $k_y$ and giving information of the edge state at that momentum.  

\begin{figure}[ht]
	\subfigure[]{ \includegraphics[width=0.45\textwidth]{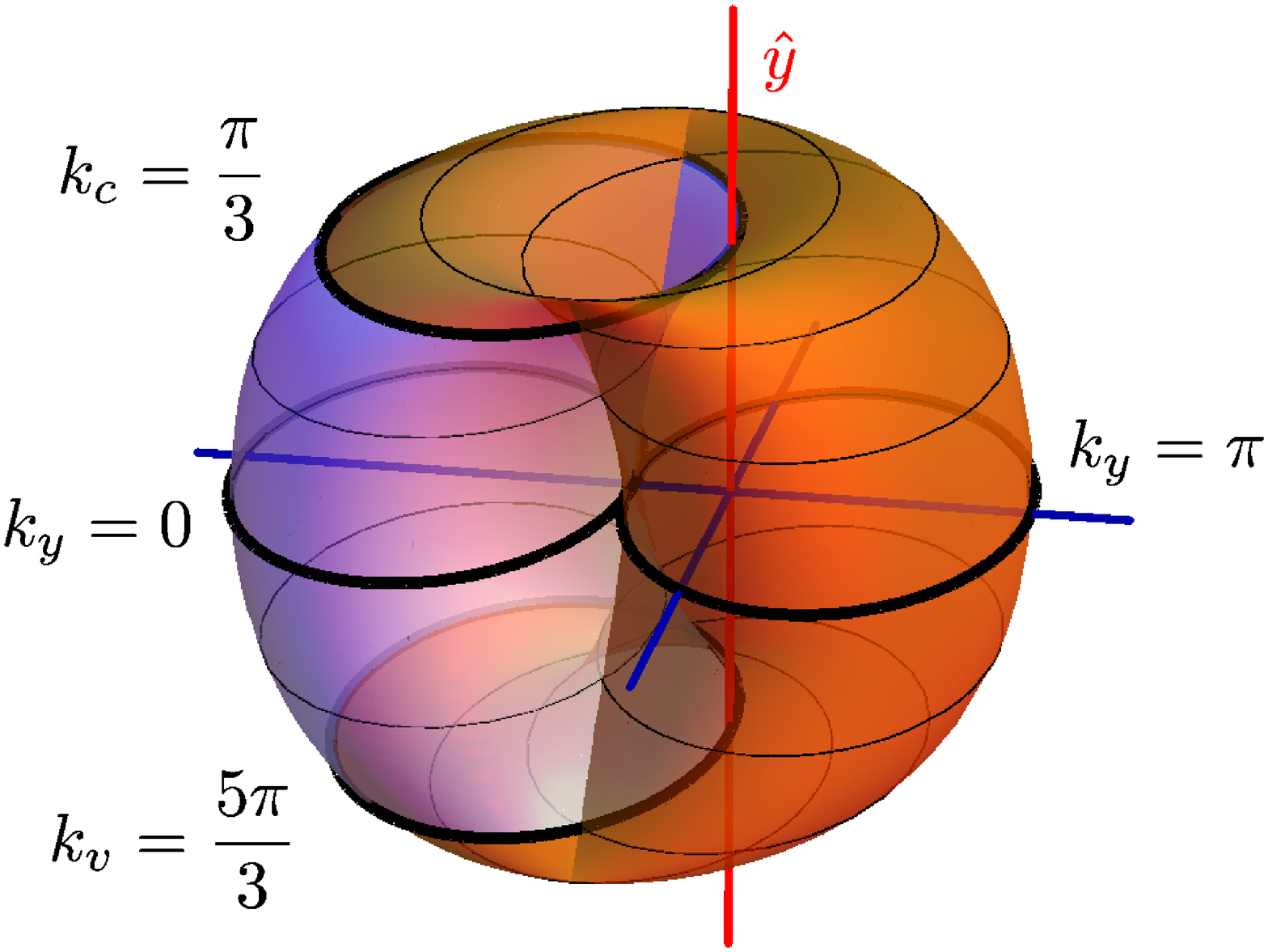}
		\label{fig:BulkBoundarya} }
	\subfigure[]{ \includegraphics[width=0.35\textwidth]{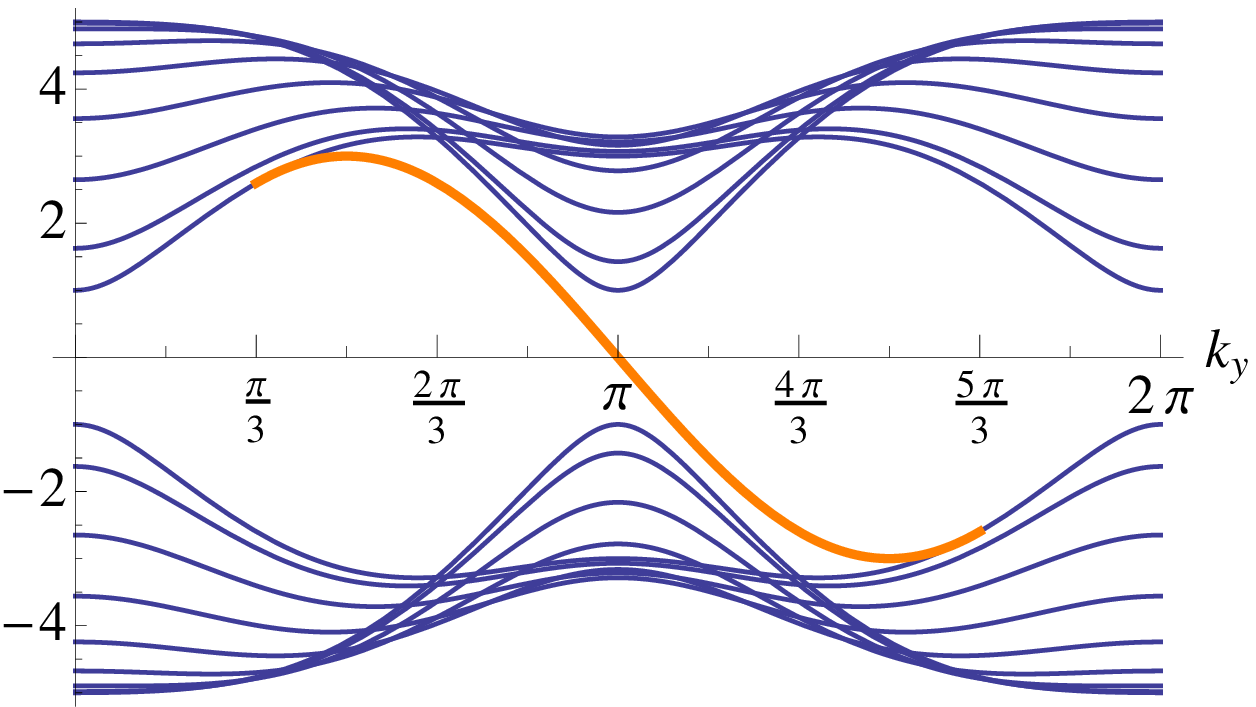}
		\label{fig:BulkBoundaryb} }
	\caption{%
		(Color online)
		Illustration of bulk-boundary correspondence.
		Figure~\ref{fig:BulkBoundarya} shows the torus traced out by $\vh(k_x,k_y)$ for a bulk insulator with Chern number $\nu=1$.
		Each black loop maps out $\vh(k_x)|_{k_y}$ for fixed values of $k_y$, the thick black lines guiding the eye to important loops.
		Setting $k_y = \pi$ gives the black loop on the right that encloses the origin, meaning there are zero-energy edge modes at this value of $k_y$.  
		At $k_y = 0$, the black loop on the left lies in the plane of the origin without containing it, indicating no edge mode at $k_y = 0$.  
		The black loops on the top and bottom ($k_v=\frac{5\pi}{3}, k_c=\frac{\pi}{3}$) have projections which intersect the origin, indicating the values of $k_y$ where the edge band merges with the bulk bands.  
		Figure~\ref{fig:BulkBoundaryb} shows the band structure of the system with the edge mode drawn in orange.
		The model presented here is a $p+ip$ superconductor described in section~\ref{sec:LatticeExamplep+ip} [see Eq.~\eqref{eq:Latticevhp+ip}] with parameters: $t = 1, \Delta_0 = 3, \mu = 1$.
	}
	\label{fig:BulkBoundary}
\end{figure}
Before proceeding to the technical proof, we present a geometric argument with the aid of Fig.~\ref{fig:BulkBoundary}, which shows an example of a bulk insulator with Chern number $\nu=1$.
The important loops of fixed $k_y$ are highlighted in black.
Since $\nu$ is nonzero and the torus wraps the origin, it is always possible to find two loops that are coplanar with the origin, one of which encloses the origin and one that does not.  
In this example, the latter case occurs at $k_y = 0$, indicating no mid-gap edge states at this $k_y$.
As we scan through different values of $k_y$, the loop moves out of this plane.
At some critical momentum $k_c$ (given by $\frac{\pi}{3}$ in Fig.~\ref{fig:BulkBoundary}), the projection of the loop onto this plane intersects the origin and an edge state emerges from the bulk conduction bands.  
At $k_y = \pi$, the loop is coplanar with the origin and encloses the origin, indicating zero-energy edge states at this value of $k_y$.  
As the plane of the loop passes through the origin, the energy of the edge state changes sign.  
The presence of edge modes for this range of momentum is shown as orange shading in Fig.~\ref{fig:BulkBoundary}.
Eventually at some critical momentum $k_v$ (given here by $\frac{5\pi}{3}$), the loop moves away from the origin and the edge state disappears in to the bulk valence band. 
Since the edge state energies at $k_v$ and $k_c$ have opposite signs, the edge band connects the bulk valence and the bulk conduction bands.

Formally, we can describe each loop $\vh(k_x)\vert_{k_y}$ by the Berry phase $\phi(k_y)$ living in a circle $[0,2\pi]$ with $0 \sim 2\pi$.~\cite{Berry}
The Berry phase can be formulated in various ways:
\begin{subequations}
\begin{align}
	\phi(k_y)
		& = -\oint_0^{2\pi}\!\! dk_x \, A_x(k_x,k_y)
			\label{eq:BPphiA}
	\\	& = \int_0^{k_y}\!\! dk_y' \oint_0^{2\pi}\!\! dk_x \, F(k_x, k_y')
			\label{eq:BPphiF}
	\\	& = \frac{1}{2} \Omega(\vh) ,
			\label{eq:BPphiSolidAngle}
\end{align}
\end{subequations}
where $A_j(\vk) = i\braket{\psi_\vk}{ \partial_j \psi_\vk}$ is the Berry connection of the filled energy states of $H(\vk)$, $F = \partial_x A_y - \partial_y A_x$ is the Berry curvature.  Geometrically, $\phi$ is half the oriented solid angle $\Omega(\vh)$ subtended by the loop $\vh(k_x)$ as seen from the origin.
The integral of $\frac{1}{2\pi}F$ over the entire Brillouin zone gives the Chern number: $\frac{1}{2\pi}\oint_\textrm{BZ}\!F = \nu$.
Both $\phi$ and $k_y$ live on a circle, and from Eq.~\eqref{eq:BPphiF}, $\phi(k_y)$ is a map $S^1 \rightarrow S^1$ with winding number $\nu$.

At the values of $\phi(k_y) = 0$ or $\pi$, the origin is in the plane of the ellipse $\vh(k_x)\vert_{k_y}$, and lies outside or inside the ellipse respectively.
Hence there is a zero energy edge state when $\phi(k_y) = \pi$, and no edge state if $\phi(k_y) = 0$ (or $2\pi$).
The family of loops as $k_y$ is varied connects these two special cases smoothly.  
For example, the upper critical momentum $k_c$ has Berry phase $0 \leq \phi(k_c) < \pi$, while the lower critical momentum $k_v$ has Berry phase $\pi < \phi(k_v) \leq 2\pi$.
Thm.~1b says that if an edge state exists, $0 < \phi < \pi$ implies it has energy $E_s > 0$, and $\pi < \phi < 2\pi$ implies $E_s < 0$.
Therefore in between $k_c < k_y < k_v$, a gapless (counterclockwise) chiral mode must exist connecting the bulk bands. 

For an insulator with Chern number $\nu$, the Berry phase $\phi(k_y)$ goes from $0$ to $2\pi\nu$ as $k_y$ is varied from $0$ to $2\pi$.
Each time the phase $\phi(k_y)$ winds around the circle, $2\pi\alpha \rightarrow 2\pi(\alpha+1)$, a pair of critical momenta $(k_{c\alpha},k_{v\alpha})$ defines a range in which a chiral mode connects the bulk valence and conduction band, $k_{c\alpha} < k_y < k_{v\alpha}$.
This chiral mode is counterclockwise as the phase $\phi$ increases by $2\pi$.
Similarly, there is a clockwise chiral mode as $\phi$ decreases by $2\pi$.
Therefore, the total number of chiral edge modes (counterclockwise $-$ clockwise) is given by the Chern number of the bulk Hamiltonian.

\subsection{Discussion}

Theorem~1 gives a simple way to compute the spectrum of edge states from properties of the bulk Hamiltonian.
The existence of zero-energy edge states is determined by whether or not the ellipse traced by $\vhpara$ encloses the origin.    
Intuitively, the size of the ellipse is a measure of the coupling strength $\vb$ between neighboring layers, while the in-plane displacement of the ellipse $\vbpara$ is a measure of coupling within the layers.
From this perspective, Thm.~1a says that an edge state exists if the nearest-layer coupling is `stronger' than the intra-layer coupling.
This is a straightforward extension of the edge states of polyacetylene, a 1D chain with alternating bond strengths $t \neq t'$, which supports an edge state if the chain terminates on the weaker bond.~\cite{SchriefferPolyacetylene79}

\begin{figure}[tb]
	\includegraphics[width=0.4\textwidth]{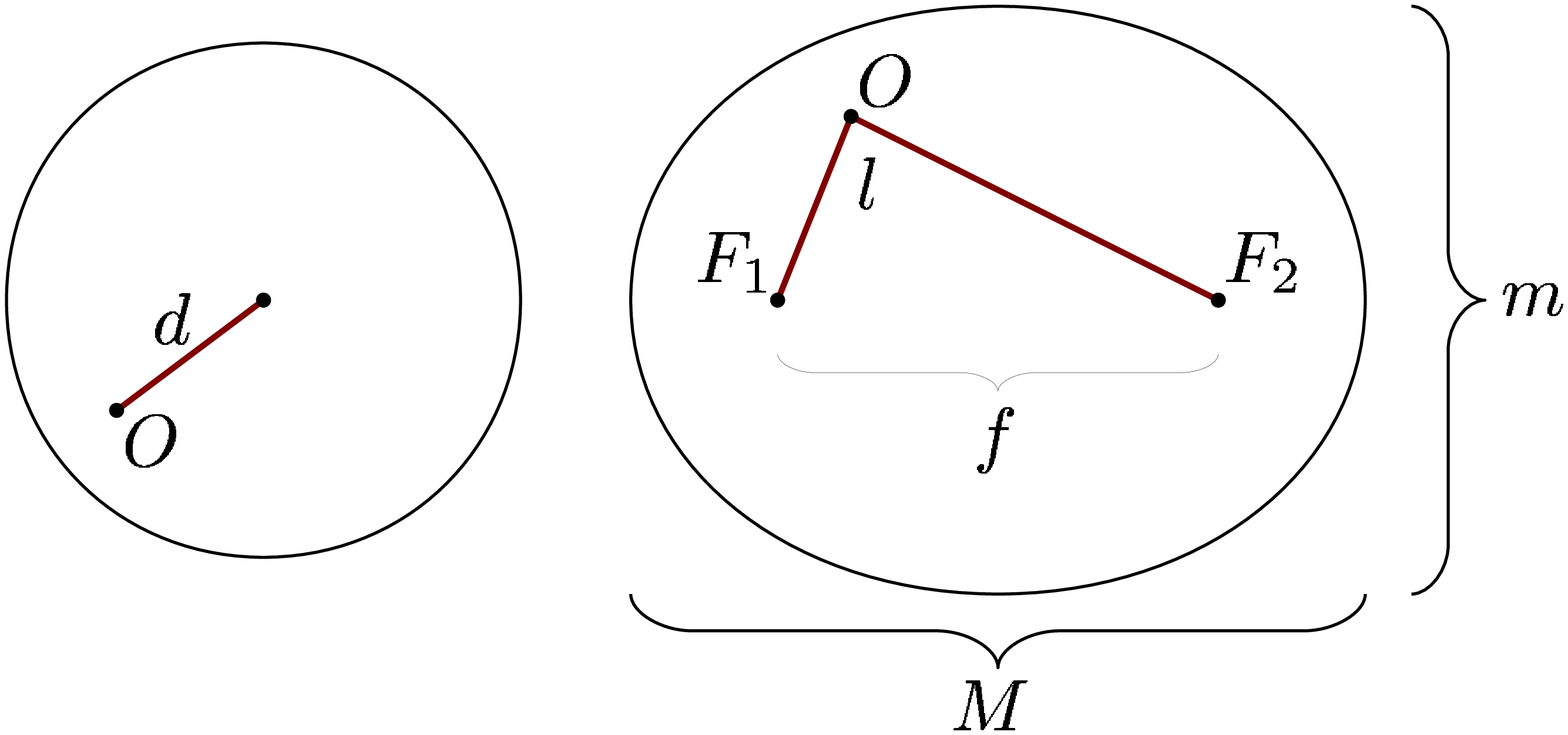}
	\caption{Determining the penetration depth from the ellipse $\vhpara$.
		The distances from the foci of the ellipse $\vhpara(\kperp)$ to the origin determine the characteristic decay parameter $\lambda$, which in turns gives the penetration depth $\xi = -a/2\ln|\lambda|$.
		In the case where $\vhpara$ traces a circle, $|\lambda| = d/r$, where $d$ is the distance of the origin to the center of the circle, and $r$ is the radius of the circle.
		In the general case $\vhpara$ traces an ellipse, $|\lambda| = \frac{l + \sqrt{l^2 - f^2}}{M + m}$, where
		$M$ and $m$ are the major and minor diameters,
		$f$ is the distance between the foci $|F_1F_2|$,
		and $l$ is the sum of distances $|OF_1| + |OF_2|$.
	}
	\label{fig:EllipseLambda}
\end{figure}

The argument presented above can also be used to calculate the  penetration depth $\xi$ of the surface states:
\begin{align}
	\xi = \frac{a}{ 2\ln(1/|\lambda|) } .
\end{align}
$a$ is the distance between layers, $\lambda$ is the characteristic decay parameter such that the wavefunction decays as $\psi_n \sim \lambda^n$ in the bulk.
$|\lambda|$ is the larger of $|\lambda_a|,|\lambda_b|$ [defined in Eq.~\eqref{eq:TMWavefunction}].
$|\lambda|$ is always less than one and is determined by the location of the origin inside the ellipse $\vhpara(\kperp)$.  
When the origin touches the edge of the ellipse, $\lambda$ has unit modulus and $\xi$ tends to infinity, indicating a bulk propagating mode.  At this point the surface spectrum ends and merges with the bulk bands.
The decay parameter can be computed by
\begin{align}
	|\lambda| = \frac{l + \sqrt{l^2 - f^2}}{M + m} ,
\end{align}
where
$M$ and $m$ are the major and minor diameters of $\vhpara(\kperp)$ respectively,
$f = \sqrt{M^2-m^2}$ is the distance between the foci of the ellipse,
and $l$ is the sum of the distances from the origin to the two foci.
(See Fig.~\ref{fig:EllipseLambda}.)

The bulk-boundary correspondence described in Thm.~2 holds even in the presence of weak interactions.
The number of edge states is a topological property of the bulk and is robust to
small perturbations that do not close the bulk insulating gap.
Such weak interactions have the effect of altering the edge state dispersion from the simple formula derived in Thm.~1.
Nevertheless, we may still view the result of Thm.~1 as a starting point where the weak interactions are added perturbatively.
With strong interactions, the electronic excitations may no longer behave as a Fermi liquid, leading to a break down of the bulk-boundary correspondence described here.
For example, electron-electron interactions may drive the 1D edge states of QSH insulators to become Luttinger liquids, which have been shown to break the topological distinction between the insulating and helical edge states.~\cite{XuMooreQSHStability06,WuBernevigHelicalLiquid06}

\section{Applications of Theorem 1}
\label{sec:LatticeExamples}

In this section, we illustrate how Theorem 1 may be used the edge state dispersion in various systems.

\subsection{Example: graphene}
\label{sec:LatticeExampleGraphene}

As an illustration of how Thm.~1 can be used to calculate edge state energies, we examine the zigzag edge of graphene.  Because the bulk energy bands are degenerate at two inequivalent points in the Brillouin zone, there is no well-defined topological invariant and Thm.~2 does not apply. 

Consider the tight-binding model for graphene on a honeycomb lattice with primitive translational vectors $\va_1$ and $\va_2$ taken to be $120^\circ$ apart.
The zigzag boundary parallel to either $\va_1$ or $\va_2$ is known to support edge modes, while the armchair boundary parallel to $\va_1- \va_2$ has no edge modes. 
Using the reciprocal space coordinates $k_1 = \vk\cdot\va_1, k_2 = \vk\cdot\va_2$, the tight-binding Hamiltonian takes the form
\begin{align}
	H & = t\MxTwo{0 & 1+e^{ik_1}+e^{-ik_2}}{1+e^{-ik_1}+e^{ik_2} & 0} ,
\end{align}
written in the $(A,B)$ basis, where $A$ and $B$ are the inequivalent sublattices.
The Hamiltonian can be written in terms of the Pauli matrices as $H = \vh \cdot \bm\sigma$ with 
\begin{align}
	\vh(\vk) = t(1 + \cos k_1 + \cos k_2, -\sin k_1 + \sin k_2, 0) .
	\label{eq:hGraphene}
\end{align}
For a zigzag edge parallel to $\va_1$, we examine the curves traced by $\vh$ for fixed $k_1$.~%
\footnote{Despite our notation, the basis vector $\va$ corresponding to $\kperp$ doesn't have to be perpendicular to the surface, it just has to be a displacement that takes one layer parallel to the surface to the next.}
The $k_2$-dependent terms in $\vh$ are written as 
\begin{align*}
	2\vb ^r \cos k_2 &= (t, 0, 0) \cos k_2 ,	\\
	2\vb ^i \sin k_2 &= (0, t, 0) \sin k_2 ,
\end{align*}
while the remaining $k_2$-independent terms become
\begin{align*}
	\vbpara &= (t + t \cos k_1, -t\sin k_1, 0) ,
	\quad\; \vbperp = 0 .
\end{align*}
$\vhpara$ lies in the plane spanned by $\vb ^r$ and $\vb ^i$, and for this system is a circle in the $xy$-plane.  It encloses the origin only if the magnitude of $\vb^0$ is less than $t$, \textit{i.e.} when $|1 + e^{ik_1}| < 1$ or equivalently when $2(1+\cos k_1) < 1$.  Thus for $\frac{2\pi}{3} < k_1 < \frac{4\pi}{3}$, $\vhpara$ encloses the origin and the system has mid-gap edge modes.
According to Thm.~1b, the energies of these edge modes are given by $\pm |\vbperp|$, which is zero.  We have shown that the zigzag boundary supports zero energy edge states when $\frac{2\pi}{3} < k_1 < \frac{4\pi}{3}$, in agreement with previous work.~\cite{FujitaGrapheneEdge96,AkhmerovGrapheneEdge08}

To see why such edge modes do not exist at armchair boundaries, consider the edge parallel to $\va_1 - \va_2$.  By rewriting the Hamiltonian in terms of $k_\parallel \equiv k_1 - k_2$ and $\kperp = k_1$, we can again examine the behavior of $\vhpara$ at fixed $k_\parallel$.  In this case, it is possible to show that $\vhpara$ never encloses the origin, and thus never satisfies the condition for mid-gap edge states.

\subsection{Example: $p+ip$ superconductor}
\label{sec:LatticeExamplep+ip}

We study a $p+ip$ system, which in the weak-pairing phase is characterized by chiral Majorana modes at the edge.~\cite{ReadGreenp+ipFQHE}
We consider a model on a square lattice with $p$-wave pairing between nonrelativistic spinless electrons, shown in Fig.~\ref{fig:BulkBoundary}.
The bulk Hamiltonian is given by
\begin{align}
	H(\vk) = \MxTwo{ \xi_\vk - \mu & \Delta_\vk }{ \Delta_\vk^\ast & -\xi_\vk + \mu } ,
	\label{eq:LatticeHamp+ip}
\end{align}
where $H$ is written in the $( c_{\vk}, c^{\dagger} _{-\vk} )^T$ basis.  The kinetic energy is given by $\xi_\vk = -2t(\cos k_x + \cos k_y)$ and the pair potential is given by $\Delta_\vk = \Delta_0 (\sin k_y + i\sin k_x)$, with $\Delta_0$ real.
Eq.~\eqref{eq:LatticeHamp+ip} can be expressed as $H = \vh \cdot \bm\tau$,
where $\tau^i$ are the Pauli matrices in Bogoliubov-de Gennes (BdG) space:
\begin{multline}
	\vh(k_x, k_y)	\\
		= (\Delta_0 \sin k_y, -\Delta_0 \sin k_x, -2t\cos k_x -2t \cos k_y -\mu) .
	\label{eq:Latticevhp+ip}
\end{multline}
For a system with edges parallel to $\hat{x}$, we examine the ellipses traced by $\vh(k_y)$ for fixed values of $k_x$.  Following Eq.~\eqref{eq:LatticehLoop}, the $k_y$-dependence of $\vh$ is decomposed as 
\begin{align*}
	2\vb^r \cos k_y &= (0, 0, -2t) \cos k_y		,	\notag\\
	2\vb^i \sin k_y &= (\Delta_0, 0, 0) \sin k_y	.
\end{align*}
The remaining $k_y$-independent term is decomposed as
\begin{align*}
	\vbpara &= (0, 0, -2t\cos k_x - \mu)		,	\notag\\
	\vbperp &= (0, -\Delta_0 \sin k_x, 0)	.
\end{align*}
For this model $\vhpara(k_y)$, which lies the plane spanned by $\vb ^r$ and $\vb ^i$, is an ellipse in the $xz$-plane.  The condition for the ellipse to enclose the origin is $|2t\cos k_x + \mu| < |2t|$, which is only possible for the range of chemical potential $|\mu| < 4|t|$.
According to Theorem~1, when this condition is satisfied, the system has mid-gap states at the left edge with energy
\begin{align*}
	E_s & = \vbperp \cdot \frac{ \vb^r\times\vb^i }{ |\vb^r\times\vb^i| }
		= \vbperp\cdot(-\hat{y})
		\\	& = \Delta_0 \sin k_x ,
\end{align*}
assuming $t\Delta_0 > 0$.
The right edge state energy is given by $-E_s = -\Delta_0 \sin k_x$.
The two edge states can become degenerate at $E_s = 0$ at either $k_x = 0$ or $\pi$ for an appropriate range of $\mu$: the degeneracy occurs at $k_x=0$ when $-4 < \mu/t < 0$, and at $k_x=\pi$ when $0 < \mu/t < 4$.  The bulk gap closes and there is a transition at $\mu/t = 0$ or $\pm4$.

\subsection{Example: 3D topological insulator}
\label{sec:LatticeExampleCubicTI}

In this section we show that the surface states of a strong topological insulator have an odd number of Dirac cones, and derive an effective surface theory.  
As an example of a topological insulator, we use a model on a cubic lattice:~\cite{HosurRyuChiralTISC}
\begin{align}
	H & = v \tau^z \Big( \sum_\mu \sigma^\mu \sin k_\mu \Big)
		+ \Big( M - t \sum_\mu \cos k_\mu \Big) \tau^x
		\label{eq:ExampleTIHam}
\end{align}
for $\mu = x,y,z$.
In the basis of Dirac matrices $\vGamma = (\tau^z\sigma^x, \tau^z\sigma^y, \tau^z\sigma^z, \tau^x)$, we can write $H = \vh\cdot\vGamma$, with
\begin{align}
	\vh(\vk) = (v\sin k_x, v\sin k_y, v\sin k_z, m(\vk)) ,
\end{align}
where the ``mass'' is given by $m(\vk) = M - t\sum\cos k_\mu$.
(The `fifth' Dirac matrix is $\tau^y$, but it is never used in this model.)
For the $(001)$ surface parallel to $\hat{x}$ and $\hat{y}$, we solve for the edge states as a function of $k_x, k_y$:
\begin{align*}
	\vh(k_z) \big\vert_{k_x,k_y}
		& = (v\sin k_x, v\sin k_y, 0, M - t\cos k_x - t\cos k_y)			\notag
			\\	&\quad + (0, 0, 0, -t)\cos k_z + (0, 0, v, 0)\sin k_z ,
\end{align*}
which lies on the $\mathit{34}$-plane (in $\Gamma$ space) displaced by $v(\sin k_x, \sin k_y)$ from the origin.
An edge state exists if and only if $|M - t\cos k_x - t\cos k_y| < |t|$ with Dirac cone spectrum $E_s = \pm v\sqrt{\sin^2 k_x + \sin^2 k_y}$.

At any of the surface time-reversal invariant momenta (TRIM) $(k_x,k_y)$,
the ellipse $\vhpara(k_z)|_{k_x,k_y}$ has one of its diameters stretching from $m(k_x,k_y,0)$ to $m(k_x,k_y,\pi)$ along the $\Gamma^4$ axis.
This ellipse encloses the origin if and only if the two endpoints straddles the origin;
equivalently, a Dirac cone appears at the surface TRIM if the mass in the bulk Hamiltonian changes sign: $m(k_x,k_y,0) m(k_x,k_y,\pi) < 0$.%
~\cite{TeoFuKaneTISurface08}

In this basis the time-reversal operator has the form $\Theta = -i\sigma^y\mathcal{K}$, where $\mathcal{K}$ is the complex conjugation operator.  The spatial inversion operator is $\Pi = \Gamma^4 = \tau^x$.
At the eight bulk TRIM, the Hamiltonian commutes with the inversion operator: $[H(\vk_\textrm{TRIM}), \Pi] = 0$.  In fact, $H$ is a multiple of $\Pi$:
\begin{align}
	H(\vk_\textrm{TRIM}) = m(\vk_\textrm{TRIM})\Pi .
		\label{eq:HamParityTRIM}
\end{align}
Hence, the sign of the mass determines the parity eigenvalue of the pair of filled states.
The strong $\ZZ_2$ topological invariant is the product of the parity eigenvalues at these TRIM, and hence $\nu_0 = \prod_\textrm{TRIM} \sgn m(\vk)$.~\cite{FuKaneTIInversion}
It is clear that the bulk $\ZZ_2$ invariant dictates whether there are an even or odd number of Dirac cones on the surface.%
~\cite{FuKaneMeleTI3D,TeoFuKaneTISurface08}

Generically, all TRI topological insulators written in $4\times4$ Dirac matrices take on a similar form to \eqref{eq:ExampleTIHam}, and most importantly they satisfy \eqref{eq:HamParityTRIM} at the eight bulk TRIM.  Hence it is straightforward to establish the bulk-boundary correspondence for 3D TRI topological insulators.

We now proceed to examine the surface eigenstates following the derivation from Sec.~\ref{sec:SurfaceEffHam}.
When $1 < m/t < 3$, the model is in the strong topological insulating phase with a single Dirac cone at $(k_x,k_y) = (0,0)$.
The surface Hamiltonian is of the form $H_s = (\vbperp\cdot\vGamma)P_s$ where $P_s$
is the projector of the two surface states: $P_s = \frac{1}{2}(1 - i(-\Gamma^4)\Gamma^3) = \frac{1}{2}(1 + \tau^y\sigma^z)$.
To examine the Hamiltonian in the reduced vector space, it is useful to apply a unitary transformation $\bar{H} = U H U^{-1}$ where $U = \exp(-i\frac{\pi}{4}\tau^x)$.
The projector becomes diagonal in the new basis
\begin{align*}
	\bar{P}_s = U P_s U^{-1} = \frac{1}{2}(1 + \tau^z\sigma^z) ,
\end{align*}
projecting onto the first and last row.
The ``edge energy'' term $\vbperp\cdot\vGamma$ transforms as
\begin{align*}
	\bar{V}^0_\perp = U (\vbperp\cdot\vGamma) U^{-1}
		& = -v \tau^y (\sigma^x \sin k_x + \sigma^y \sin k_y)
		\\	& \approx -v \tau^y (\sigma^x k_x + \sigma^y k_y) .
\end{align*}
Since $\bar{P}_s$ is diagonal with entries $(1,0,0,1)$, it suffices to examine only the four corners of $\bar{V}^0_\perp$.
The effective surface Hamiltonian can be computed:~%
\footnote{The Rashba coupling results from our choice of $U$.  It is also possible with a different choice of $U$ to arrive at the surface Hamiltonian of the form $\bm\sigma\cdot\vk$.}
\begin{align}
	\bar{H}_s = \bar{V}^0_\perp \bar{P}_s
		& \approx v\MxTwo{0 & k_y+ik_x}{k_y-ik_x & 0}			\notag
		\\	& = v (\bm\sigma \times \vk) \cdot \hat{z} .
\end{align}
The basis of the surface Hamiltonian $\bar{H}_s$ is, in terms of the basis of $H$, $(1,0,i,0)^T/\sqrt{2}$ and $(0,1,0,i)^T/\sqrt{2}$ (The first and last columns of $U^\dag$).  The spin degrees of freedom ($\bm\sigma$) and the orbital degree of freedom ($\bm\tau$) are entangled in the surface states.


\section{Continuum Hamiltonians quadratic in momentum}
\label{sec:ContinuumBulkEdge}

The bulk-edge correspondence stated in Sec.~\ref{sec:NNBulkEdge} may also be extended to Hamiltonians in the continuum.
Given a translationally invariant Dirac Hamiltonian quadratic in momentum $\vp = -i\nabla$, of the form:
\begin{align}
	H(p; \vp_\parallel) & = C^0(\vp_\parallel) + C^1(\vp_\parallel) p + C^2(\vp_\parallel) p^2
			\notag
	\\	& = \left[\vc^0(\vp_\parallel) + \vc^1(\vp_\parallel) p + \vc^2(\vp_\parallel) p^2\right]
				\cdot \vGamma
		\label{eq:ContinuumHamiltonian}
\end{align}
where $\vp_\parallel$ and $p$ are, respectively, the momentum parallel and perpendicular to the edge/surface.
For a fixed momentum $\vp_\parallel$, the vector
\begin{align}
	\vh(p) \big\vert_{\vp_\parallel} = \vc^0 + \vc^1 p + \vc^2 p^2
\end{align} traces a parabola.  This parabola lies on some 2D plane spanned by $\vc^1$ and $\vc^2$, and we can always decompose $\vc^0$ and $\vh$ into in-plane and out-of-plane components: $\vc^0 = \vcperp + \vc^0_\parallel$ and $\vh(p) = \vcperp + \vhpara(p)$.
In this section we state the main theorem:

\textbf{Theorem 3a.}
\textit{%
An edge state (with zero Dirichlet or Neumann boundary condition) exists if and only if the origin is within the concave side of the parabola $\vhpara(p)$.%
}

\textbf{Theorem 3b.}
\textit{%
The energy of the edge state is given by the distance of the plane to the origin, \textit{i.e.} $E_s = \pm|\vcperp|$.
When the gamma matrices are the Pauli matrices, the left edge (semi-infinite slab with $x \geq 0$) energy is given by:
$E_s = \vc^0 \cdot \frac{ \vc^1\times\vc^2 }{ |\vc^1\times\vc^2| }$.
}

The proof of Theorem~3 is given in Appendix~\ref{sec:ProofContinuumTM}.

\subsection{Discussion}

The proof of Thm.~3 uses the ansatz $\psi(x) = u_a e^{i\kappa_ax} + u_b e^{i\kappa_bx}$ and derives the condition when $\Im \kappa_{a,b} > 0$ for edge states.
Physically, $\xi = (\Im\kappa)^{-1}$ gives the penetration depth of the edge modes.
We may think of the continuum as a limiting case of the lattice as the lattice spacing $a$ goes to zero.  
The ellipse $\vhpara$ becomes a parabola for vanishing $a$ as the quantities $\vc^0 = \vb^0 + 2\vb^r$, $\vc^1 = 2\vb^i a$, $\vc^2 = -\vb^r a^2$ are held constant.
Theorem~2 also extends to the continuum case as the paraboloid $\vh(k_x,k_y)$ determines the number of chiral edge states.

In addition to Dirichlet and Neumann boundary conditions, there is also a mixed type with $\psi'(0) = \eta\psi(0)$ for a positive number $\eta$.  The origin of this boundary condition comes from requiring the wavefunction outside ($x<0$) to satisfy $H_\textrm{vac} = W + \frac{p^2}{2m}$, where $W>E_s$ is the work function.  Our analysis and result holds even for this boundary condition.

Notice that the quadratic term $C^2 p^2$ is crucial for the existence of edge states.
Without it, the polynomial \eqref{eq:ContinuumQuartic} will be quadratic and there can only be one solution for $\kappa$ in the upper half plane.  The form of the solution $\psi = ue^{i\kappa x}$ makes it impossible to satisfy either type of boundary condition.

In the lattice model, we can compute the edge spectrum only for certain surfaces because of the nearest\hyp{}layer requirement.  For example, we can only compute the $\{100\}$, $\{110\}$ and $\{111\}$ surface dispersion of the TI model on a cubic lattice.
In the continuum case with the Hamiltonian bilinear in momentum, any surface cut will still yield a Hamiltonian quadratic in $p_\perp$.
Consequently, we can compute the surface excitation spectrum and wavefunctions of the system for all linear surfaces.


\subsection{Example: $p+ip$ superconductor}

We use the simplest model of a $p+ip$ superconductor:~\cite{ReadGreenp+ipFQHE}
\begin{align}
	H(p_x,p_y) & = \MxTwo{ \frac{p^2}{2m^\ast}-\mu & \Delta_0(p_x-ip_y) }
			{ \Delta_0(p_x+ip_y) & \mu-\frac{p^2}{2m^\ast}} .
\end{align}
Equivalently, with $H(\vp) = \vh(\vp) \cdot \bm\tau$,
\begin{align}
	\vh(\vp) & = \left( \Delta_0 p_x, \Delta_0 p_y, \frac{p^2}{2m^\ast}-\mu \right) .
\end{align}
This model is isotropic, and without loss of generality, we take a semi-infinite plane $x\geq0$ with $\hat{y}$ parallel to the edge.
At a fixed $p_y$, the $\vh$ vector becomes:
\begin{align*}
	\vh(p_x) & = \left(0, \Delta_0 p_y, \tfrac{p_y^2}{2m^\ast}-\mu\right)			\notag
		\\	&\quad	+ (\Delta_0, 0, 0)p_x + \left(0, 0, \tfrac{1}{2m^\ast}\right)p_x^2
\end{align*}
which lies in the $xz$-plane with $y = \Delta_0 p_y$.  The parabola is concave towards the $+\hat{z}$ direction, and hence an edge state exists if and only if $p_y^2/2m^\ast-\mu < 0$.
Edge states can only exist when $\mu$ is positive, or in other words in the `weak pairing phase' of $p+ip$ superconductors.

The edge state energy dispersion is given by Thm.~3b.
\begin{align*}
	E_s & = \vc^0 \cdot \frac{ \vc^1\times\vc^2 }{ |\vc^1\times\vc^2| }			\notag
			= \Delta_0 p_y \hat{y} \cdot (\hat{x} \times \hat{z})			\notag
		\\	& = -\Delta_0 p_y	.
\end{align*}
Referring to Eq.~\eqref{eq:ProjectorEdgeStates},
the edge state wavefunction is given by the projector $P_s = \frac{1}{2}(1 - i\tau^x\tau^z) = \frac{1}{2}(1 - \tau^y)$.
Hence the edge states parallel to the $y$-axis are eigenstates of $\tau^y$.


\section{Outlook}

In this paper, we provided two main results.
Theorem~1 gives a general prescription for finding edge states of Dirac Hamiltonians (with nearest-layer coupling) on a lattice.
Its range of applicability includes ``accidental edge states'' which may not be topologically protected, such as in boron nitride.
Theorem~2 relates the bulk Chern number of a 2D insulator to the number of chiral edge modes.  This establishes the bulk-boundary correspondence for a class of quantum Hall insulators.

For 3D time-reversal invariant insulators, we demonstrated in Sec.~\ref{sec:LatticeExampleCubicTI} how the $\ZZ_2$ strong topological invariant determines whether there is an odd or even number of Dirac cones in the surface spectrum.
Although we have used a specific TI model in the example, the argument is easily generalizable for all TRI Dirac Hamiltonians.

This work can be extended beyond quantum Hall insulators (class A) and TRI topological insulators (class AII) to other insulators within the Altand and Zirnbauer classification.~\cite{AltlandZirnbauer97,ZirnbauerDisorderedFermions05}
The periodic table of topological insulators and superconductors provide an exhaustive topological classification of non-interacting electronic systems.~\cite{SFRLClassification3D,KitaevClassification,SFRLClassificationSurface}
As there are model Dirac Hamiltonians~\cite{HaldaneQHE88,FuKaneMeleTI3D,HosurRyuChiralTISC,RyuLudwigDimHeir10} in each class, our work provides the machinery to relate the bulk topological invariants~\cite{KitaevClassification} to the surface properties~\cite{SFRLClassification3D,SFRLClassificationSurface} of these systems.

\begin{acknowledgments}

The authors thank D.-H.~Lee for suggesting the approach using Green's functions.  
We gratefully acknowledge discussions with A.~M.~Essin and P.~Ghaemi, and useful comments and feedback on the manuscript from J.~H.~Bardarson, J.~Cayssol, J.~E.~Moore,  and M.~Zaletel.

This work is supported by NSF DMR-0804413 (VS) and the Center for Functional Engineered Nanoarchitectures (RM).

\end{acknowledgments}

\appendix

\clearpage
\section{Proof by Green's functions}
\label{sec:ProofGF}

\newcommand{\vhparamin}{ {\vh_\parallel^\textrm{min}} }

It is difficult to study a system with edges because of the broken translational invariance. 
We begin by writing the Green's function for an easier problem: a periodic system with no boundary.~\cite{ButtnerShockleyStates72}
The full translational invariance allows us to work in momentum space, reducing the dimension of the Hamiltonian in momentum space.  
Next, the geometry of the system is changed from a periodic to an open system by subtracting all interactions between a particular pair of nearest neighbors.  We use the Dyson equation to calculate the Green's function for the open geometry and show that there are poles -- and thus bound states -- at mid-gap energies.  The form of the potential required to cut the periodic system greatly reduces the degrees of freedom in the problem and enables an analytic solution.

We prove Theorem~1 for the case of $2\times 2$ Dirac Hamiltonians, which can be decomposed in terms of the Pauli matrices $\sigma ^i$.  
For higher dimensional Hamiltonians, it is always possible to find three gamma matrices $\Gamma ^i$ that whose sub-blocks are the Pauli matrices, \textit{e.g.} $\tau^z \sigma^x, \tau^z \sigma^y, \tau^z \sigma^z$.
A suitable unitary transformation rotates the Hamiltonian so that it is a linear combination of these three gamma matrices, and the arguments of this section apply to each sub-block.   

\subsection{Bulk Green's function}
We prove Theorem 1 for $2\times 2$ Hamiltonians, which can be expressed in the following form:

\begin{equation}
	H(k) = E_k \begin{bmatrix}
		\cos \theta	&	\sin \theta \, e^{-i\phi}  \\
		\sin \theta \, e^{i\phi}	&	-\cos \theta
	\end{bmatrix}
	\;.
	\label{ham2d}
\end{equation}  
Decomposing $H$ into the Pauli matrices $\sigma^i$ gives $H = \vh \cdot \bm\sigma$ for
\begin{align}
	\vh = E_k( \sin \theta \cos \phi, \sin \theta \sin \phi, \cos \theta )	.
	\label{hproof}
\end{align}
The eigenstates are given by
\begin{align}
	\ket{\psi_-}
		&= \begin{pmatrix} \sin\frac{\theta}{2} \\ -\cos\frac{\theta}{2} e^{i\phi} \end{pmatrix} ,	&
	\ket{\psi_+}
		&= \begin{pmatrix} \cos\frac{\theta}{2} \\ \sin\frac{\theta}{2} e^{i\phi} \end{pmatrix} .
	\label{eigenstates}
\end{align}
To express $\vh$ in terms of the parameters defined in Section~\ref{sec:NNBulkEdge}, we assume $\vb ^r$ and $\vb ^i$ lie along $+\hat{x}$ and $+\hat{y}$ respectively, with $\vbperp$ along $\hat{z}$.  The Hamiltonian takes the form
\begin{align}\begin{split}
	E_k \sin \theta \cos \phi \,\hat{x}
		&= 2\vb ^r \cos k + (\vbpara \cdot \hat{x}) \,\hat{x} ,  \\
	E_k \sin \theta \sin \phi \,\hat{y}
		&= 2\vb ^i \sin k + (\vbpara \cdot \hat{y}) \,\hat{y} ,	\\
	E_k \cos \theta \,\hat{z}  &= \vbperp .
	\label{matrixelemdef}
\end{split}\end{align}
Here $\vhpara$ lies in the $xy$-plane.  
For the remainder of this section, we will use the notation $b_r = |\vb ^r|$ and $b_i = |\vb ^i|$.  
Note that any Hamiltonian may be brought to this form by an appropriate gauge transformation.

The first step in calculating the edge mode energies is to write the Green's function for the  fully periodic system.  Utilizing the full translational invariance, we work for now in the momentum representation.  The $\alpha\mbox{\hyp{}}\beta$ matrix element of the bulk Green's function is given by
\begin{align}
	 G_0 ^{\alpha \beta}\left( E; k \right)
		= \sum_i {\frac{ \ket{\psi_i}^\alpha \bra{\psi_i}^\beta }{E - E_i}}
	\;,
	\label{defgreen}
\end{align}
where $i$ sums over the energy eigenstates of $H_\vk$.  The four matrix elements are given by
\begin{subequations}
	\label{green2d}
\begin{align}
	G_0 ^{11} \left( E; k \right) &= \frac{E + E_k \cos \theta}{E^2 - E_k ^2}
		\label{green2d_11}	,\\
	G_0 ^{22} \left( E; k \right) &= \frac{E - E_k \cos \theta}{E^2 - E_k ^2}
		\label{green2d_22}	,\\
	G_0 ^{12} \left( E; k \right) &= E_k\frac{\sin \theta \cos \phi - i\sin \theta \sin \phi}{E^2 - E_k ^2}
		\label{green2d_12}	,\\
	G_0 ^{21} \left( E; k \right) &= E_k\frac{\sin \theta \cos \phi + i\sin \theta \sin \phi}{E^2 - E_k ^2}
		\label{green2d_21}	.
\end{align}
\end{subequations}

In order to include the effects of a boundary that is localized in real space, the bulk Green's function is written in real space via a Fourier transformation.  
\begin{align}
	G_0(E; y) = \begin{bmatrix}
		B^0 & B^{\dagger} & \makebox[0mm]{$\hspace{1mm}\cdots$} & & B \\
		B & B^0 & B^{\dagger} & &\\
		\raisebox{-1mm}[0mm][0mm]{\vdots} & B & B^0 & B^{\dagger}&\\
		&&B & B^0 &\\
		B^{\dagger}&&&&
			\makebox[6mm]{\hspace{-4mm} \raisebox{2mm}[6mm]{$\ddots$} }
	\end{bmatrix}
	\label{realspaceG}
\end{align}
for 
\begin{align}
	B^0 = \int \frac{dk}{2\pi} G_0(E; k), \ 
	B = \int \frac{dk}{2\pi} e^{ik}G_0(E; k)
	.	\label{def_fourier}
\end{align} 
There are of course more non-zero matrix elements, corresponding to the mixing of matrix elements separated by more than one lattice constant.  However, for systems with nearest-neighbor interaction, the matrix elements contained in $B^0$ and $B$ are the only ones needed to prove the existence of zero energy states.

\subsection{Green's function of the open system}
Next we write an expression $V$ that deletes the coupling terms to create a system with edges.  In a system with nearest-layer  interactions, the only non-zero matrix elements are those between a single pair of neighboring layers:
\begin{equation}
	V = \begin{bmatrix}
		0 & \cdots & & V_b \\
		{\vdots} & 0 &&\\
		&& \ddots &\\
		V_b^\dagger &&& 0
	\end{bmatrix} \;.
	\label{V2d}
\end{equation}
$V_b = -\vb \cdot \bm\sigma$, and for the Hamiltonian described in Eq.~\eqref{matrixelemdef} takes the form
\begin{equation}
	V_b = \begin{bmatrix}
		0 & -b_r - b_i\\
		-b_r + b_i & 0 \end{bmatrix} \;.
	\label{Vb}
\end{equation}
 
We are now ready to compute the Green's function for the open system.  The Dyson equation gives an exact expression for the open Green's function $G$ in terms of the periodic Green's function $G_0$ and the cuts $V$ needed to take the system from one geometry to the other:
\begin{equation}
	G(E; y) = \big( I - G_0(E; y)V \big)^{-1} G_0(E; y) .
	\label{dysonApp}
\end{equation}
Substituting Eq.~\eqref{V2d} and \eqref{realspaceG} into Eq.~\eqref{dysonApp} gives the following condition for an edge state wavevector $u$:
\begin{multline}
	(1 - G_0 V) u
	\\	= \begin{bmatrix}
		I - B V_b ^{\dagger} && &&& -B^0 V_b \\
		& I &&&&\\
		\vdots && \ddots &&&\vdots \\
		&&&& I &\\
		- B^0 V_b ^{\dagger} && &&& I - B  ^{\dagger} V_b \end{bmatrix} u = 0 .
	\label{bigmatrix}
\end{multline}
The ellipses indicate the only non-zero sub-blocks: the first and last columns, and copies of the identity along the diagonal.

\subsection{Existence and spectrum of edge modes}
Since an edge state wavevector $u$ has non-zero components in the sub-block corresponding to one of the edges, Eq.~\eqref{bigmatrix} can be satisfied in two ways. 
There is an edge mode on the left edge when the first two columns of Eq.~\eqref{bigmatrix} are linearly dependent, and on the right edge when the last two columns are linearly dependent.  
Recalling the expressions for the bulk Green's functions, Eq.~\eqref{green2d}, we introduce some notation for the open Green's function:
\begin{subequations}
	\label {matelema}\
\begin{align}
	a_0 &= \int \frac{dk}{2\pi} \frac{E}{E^2 - E_k ^2}
	\label {matelema0},\\
	a_x &=  \int \frac{dk}{2\pi} \frac{E_k \sin \theta \cos \phi}{E^2 - E_k ^2}
	\label {matelemax},\\
	a_y &=  \int \frac{dk}{2\pi} \frac{E_k \sin \theta \sin \phi}{E^2 - E_k ^2}
	\label {matelemay},\\
	a_z &=  \int \frac{dk}{2\pi} \frac{E_k \cos \theta}{E^2 - E_k ^2}
	\label {matelemaz}.
\end{align}
\end{subequations}
$c_0 = \int\! \frac{dk}{2\pi} e^{ik}\frac{E}{E^2 - E_k ^2}$, and similar for $c_x, c_y$, and $c_z$.  
Let us first examine the conditions required for a left edge mode.  Collapsing Eq.~\eqref{bigmatrix} to the subspace corresponding to the two edges, the first two columns can be written as 
\begin{align}
	L_1 &= \begin{bmatrix}
		1 + (b_r + b_i)(c_x - ic_y)\\
		(b_r + b_i)(c_0 - c_z)\\
		(b_r + b_i)(a_x - ia_y)\\
		(b_r + b_i)(a_0 - a_z)\end{bmatrix}
		\;,	\\
	L_2 &= \begin{bmatrix}
		(b_r - b_i)(c_0 + c_z)\\
		1 + (b_r - b_i)(c_x + ic_y)\\
		(b_r - b_i)(a_0 + a_z)\\
		(b_r - b_i)(a_x + ia_y) \end{bmatrix} \;.
\label{left_const}
\end{align}

Now we examine the conditions required for the system to have a left edge state, \textit{i.e.} when the columns $L_1$ and $L_2$ are linearly dependent.     
This is done by writing $L_1$ and $L_2$ as a $4\times 2$ matrix $M$ and requiring that the determinant of any $2\times 2$ sub-block vanish.  
This constraint takes on the following form for the sub-blocks of $M$ created by the bottom two rows, the top two rows, and the second and fourth rows, respectively:
\begin{subequations}
\begin{align}
	0 &= (b_r ^2 -b_i ^2)(a_x ^2 + a_y ^2 - a_0 ^2 + a_z ^2)
		\label{plus_const_1},\\
	0 &= 1 + 2(b_r c_x - ib_i c_y) \notag
		\\ &\quad + (b_r ^2 -b_i ^2)(c_x ^2 + c_y ^2 - c_0^2 + c_z ^2)
		\label{plus_const_2},\\
	0 &= (b_r ^2 -  b_i ^2)(c_0 - c_z)(a_x + ia_y)	\notag
		\\	&\quad	-(b_r ^2 -b_i ^2)(c_x +ic_y)(a_0 - a_z) -  (b_r + b_i)(a_0 - a_z)
		\label{plus_const_3}.
\end{align}
\end{subequations}
The conditions are equivalent to $B^0 V_b ^{\dagger}u_L = 0$ and $(I - B V_b ^{\dagger})u_L = 0$, which together require $\Det{B^0} = 0$.  
To see this, note that although the first condition is satisfied when either $\Det{B^0} = 0$ or $V_b ^{\dagger}u_L = 0$, the second case cannot satisfy $(I - B V_b ^{\dagger})u_L = 0$.

\subsection{Constraints on $\vhpara$ and $E^2$}
We begin by showing that
\begin{align}
	\Det{B^0} = a_0^2 - a_x^2 - a_y^2 - a_z^2
	\label{DetB0}
\end{align}
is zero if and only if $\vhpara$ encloses the origin and the edge state energy is given by $E = \pm |\vbperp|$.  
For a geometric view of the variables $a_0, a_x$, \textit{etc.}, note that $E_k (\sin \theta \cos \phi, \sin\theta \sin \phi)$ are the $(x, y)$ coordinates of $\vhpara$, and $E_k \cos \theta = \vbperp \cdot \hat{z}$.  
 $E_k^2$ is given by  $|\vbperp|^2 + |\vhpara|^2$, and $E^2 - E_k^2 < 0$ for a mid-gap state.
We examine two cases, when $\vhpara$ encloses the origin and when it does not.

Suppose the ellipse does not enclose the origin.
Let $\vhparamin$ be the point on the ellipse $\vhpara$ closest to the origin.
Because the ellipse is convex and does not contain the origin, it must lie in the half of the plane (spanned by $\vb^r,\vb^i$) for which $\vr \cdot \vhparamin$ is positive.
In other words, $\vhpara(k) \cdot \vhparamin$ is positive definite.
Moreover, $\vhpara(k) \cdot \vhparamin \geq |\vhparamin|^2$ from its definition.
The Cauchy-Schwarz inequality says that
\begin{align}
	\big| (a_x\hat{x} + a_y\hat{y}) \cdot \vhparamin \big|
		\leq | a_x\hat{x} + a_y\hat{y}  | \, \big| \vhparamin \big| ,
\end{align}
therefore
\begin{align}
	| a_x\hat{x} + a_y\hat{y} |				\notag
		& \geq \frac{1}{|\vhparamin|} \int \frac{dk}{2\pi} \frac{\vhpara \cdot \vhparamin}{E_k^2 - E^2}
	\\	& \geq \frac{1}{|\vhparamin|} \int \frac{dk}{2\pi} \frac{ |\vhparamin|^2 }{E_k^2 - E^2} .
\end{align}
This sets an upper bound to $\Det{B^0}$:
\begin{align}
	a_0 ^2 - a_z^2 - a_x ^2 - a_y ^2
		\leq \left( E^2-|\vbperp|^2 \right) \mathcal{I}^2 - |\vhparamin|^2 \mathcal{I}^2 ,
\end{align}
where $\mathcal{I} = \int\! \frac{dk}{2\pi} \frac{1}{E_k^2 - E^2}$.
Since $|\vbperp|^2 + |\vhparamin|^2 > E^2$, the expression is always negative and never zero.
Hence no edge states can exist when the ellipse $\vhpara$ fails to enclose the origin.

Now we consider the case when the ellipse $\vhpara$ encloses the origin and show that an edge state exists only when $E^2 = |\vbperp| ^2$.  
First, when $E^2 = |\vbperp|^2$, the integrals $a_x$ and $a_y$ are both zero.
Because $E = |\vbperp|$, $a_0 ^2 - a_z^2 = 0$ and the determinant is zero.

To see why $a_x = a_y = 0$ in this case, note that the denominator of $\mathcal{I}$ becomes $|\vhpara |^2$, and we can express the conditions geometrically.  
Here we use Gauss's law in 2D to show that $\int\!dk \frac{\vhpara}{|\vhpara|^2}$ is zero if and only if $\vhpara$ encloses the origin.
We can visualize the expression from an electrostatics point of view:
for a charged ellipse in 2D with a charge distribution
$\int\!dk \,\delta^2(\vr-\vhpara)$, the electric field
($\propto \frac{1}{r}$ in 2D) at the origin is given by $\int\!dk \frac{\vhpara}{|\vhpara|^2}$.
Analogous to a uniformly charged circle, the charges are distributed such that the electric field is vanishing in the
interior but nonzero in the exterior:
\begin{equation}
	a_x\hat{x} + a_y\hat{y} = -\int\!dk \frac{\vhpara}{|\vhpara|^2} = 0 .
	\label{geometric1ellipseint1}
\end{equation} 

Now we show that $\Det{B^0}=0$ is not satisfied for any other value of $E^2$.  If $E^2 < |\vbperp|^2$ then $|E| < |\vbperp| = |E_k \sin\theta \sin \phi|$.  
It follows that $a_0^2 - a_z^2 = (E^2 - |\vbperp|^2)\mathcal{I}^2 < 0$, hence there are no edge states. 

If $E^2 > |\vbperp|^2$, we let $E^2 = \epsilon^2 + |\vbperp|^2$, such that $E_k^2 - E^2 = |\vhpara|^2 - \epsilon^2 > 0$.
$a_0$ and $a_z$ can be written as:
\begin{subequations}
\begin{align}
	-a_0 &= \int \frac{dk}{2\pi} \frac{E}{|\vhpara|^2 - \epsilon^2} ,
	\\
	-a_z &=  \int \frac{dk}{2\pi} \frac{|\vbperp|}{|\vhpara|^2 - \epsilon^2} ,
\end{align}
which combine to give:
\begin{align}
	a_0^2 - a_z^2 = (|E^2 - \vbperp|^2) \mathcal{I}^2 = \epsilon^2 \mathcal{I}^2 .
\end{align}
For $a_x$ and $a_y$, we use the manipulation
\begin{align*}
	\frac{1}{|\vhpara|^2 - \epsilon^2} - \frac{1}{|\vhpara|^2}
		= \frac{\epsilon^2}{ (|\vhpara|^2 - \epsilon^2) |\vhpara|^2 }
\end{align*}
to get:
\begin{align}
	-a_x\hat{x} - a_y \hat{y} &= \epsilon \int \frac{dk}{2\pi}
		\frac{ \vhpara \epsilon }{ (|\vhpara|^2 - \epsilon^2) |\vhpara|^2 } .
\end{align}
\end{subequations}
By the triangle inequality: $\big| \int\!dk\, \mathbf{u}(k) \big| \leq \int\!dk\, |\mathbf{u}(k)|$, we can put an upper bound
\begin{align}
	\left|  a_x \hat{x} + a_y \hat{y} \right|
		& \leq \epsilon \int \frac{dk}{2\pi}			\notag
			\frac{ |\vhpara| \epsilon }{ (|\vhpara|^2 - \epsilon^2) |\vhpara|^2 }
	\\	& = \epsilon \int \frac{dk}{2\pi} \frac{ \epsilon }{ (|\vhpara|^2 - \epsilon^2) |\vhpara| }			\notag
	\\	& < \epsilon \int \frac{dk}{2\pi} \frac{ 1 }{ (|\vhpara|^2 - \epsilon^2) }			\notag
	\\	& = \epsilon \mathcal{I} .
\end{align}
We have used the fact that $|\vhpara| > \epsilon$ to go from the second to third line.
This implies $a_x ^2 + a_y ^2 < \epsilon^2\mathcal{I}^2$ and sets a lower bound to Eq.~\eqref{DetB0}: $a_0^2 - a_z^2 - a_x ^2 - a_y ^2 > 0$ for $|E| > |\vbperp|$.

Hence we have shown that $\Det{B^0} = 0$ if and only if $E^2 = |\vbperp|$ and the ellipse $\vhpara$ encloses the origin.

Turning now to the second constraint, Eq.~\eqref{plus_const_2}, we note that $E = \pm |\vbperp|$ implies that $c_0 ^2 = c_z ^2 $.  
In the following, we use the fact that $c_x = -i c_y$.  
Substituting this into Eq.~\eqref{plus_const_2} gives the constraint $1 + 2(b_r c_x - ib_i c_y) = 0$, which may be expressed as 
\begin{equation}
	1 - 2\int \frac{dk}{2\pi} e^{ik} \frac{\vhpara \cdot \vb ^*}{|\vhpara|^2} = 0 .
	\label{ellipseint2}
\end{equation}  
In order to satisfy Eq.~\eqref{plus_const_1} and Eq.~\eqref{plus_const_2} simultaneously, Eq.~\eqref{ellipseint2} must be true if and only if $\vhpara$ encloses the origin.  
This can be shown for an arbitrary ellipse centered at $\vbpara$ and with semimajor and semiminor axes given by $2\vb ^r$ and $2\vb ^i$.  

\subsection{Sign of the energy}
The final constraint, Eq.~\eqref{plus_const_3}, determines the sign of the edge state energy.  Note that the first two terms vanish because $a_x = a_y = 0$ and $c_x = -ic_y$.  
Getting rid of the remaining term requires choosing a particular sign for the energy.  
The condition is satisfied with the choice $a_0 = a_z$. 

The constraints for the left edge can be summarized as 
\begin{list}{1\alph{enumi}.}%
	{\leftmargin=3.2\parindent \topsep=1mm \parsep=0mm \parskip=0mm}
	\usecounter{enumi}
\item	$E = +\vbperp \cdot \hat{z}$,
\item	$1 + 2(b_r  c_x - i b_i c_y) = 0$.  
\end{list}

A similar calculation for the right edge, using the last two columns of Eq.~\eqref{bigmatrix}, gives the following conditions: 
\begin{list}{2\alph{enumi}.}%
	{\leftmargin=3.2\parindent \topsep=1mm \parsep=0mm \parskip=0mm}
	\usecounter{enumi}
\item	$E = -\vbperp \cdot \hat{z}$,
\item	$1 + 2(b_r c_x  + i b_i c_y )  = 0$.
\end{list}

Condition 2b. can be expressed as $1 - 2\int\! \frac{dk}{2\pi} e^{-ik} \frac{\vhpara \cdot \vb }{|\vhpara|^2} = 0$.  Hence it is also satisfied when $\vhpara$ encloses the origin, but with opposite orientation as for the left edge.  
The conditions for edge modes are now equivalent to two statements:
\begin{enumerate}
\item	The function $\vhpara$ must enclose the origin,
\item	The energy is given by $E = \pm \vbperp \cdot \hat{z}$, with the sign determined by the orientation of loop $\vhpara$.
\end{enumerate}

We have succeeded in analytically deriving the condition for the $2L\times2L$ matrix representing the Hamiltonian for an open system to have zero eigenvalues.  The power of the method lies in the fact that $V$ has non-zero matrix elements only in the $4\times4$ subspace of electron operators at the two edges.  Thus the effect of the boundary can be seen by examining the $4\times4$ subspace, which can be handled analytically.

\section{Proof by transfer matrices}
\label{sec:ProofTM}

In this section, we prove Theorem 1 for a hard-edge at a fixed $\vkpara$.
We begin by defining a new function $\vbeta$ (which is like a complex extension of $\vh$), and the form of our edge states.

In the Hamiltonian \eqref{eq:LatticeHopping},
$\vb\cdot\vGamma,\vb^0\cdot\vGamma,\vb^\ast\cdot\vGamma$ are the hopping matrices from the previous, same, and next layers respectively.
We consider a semi-infinite system where the layers are labelled by positive integers with $n=1$ denoting the layer at the surface.  
Hence we ignore the terms $\Psi_1^\dag \Psi_0^\phd$ and $\Psi_0^\dag \Psi_1^\phd$ in considering our semi-infinite system.
Consequently, an excitation $\psi^\dag = \sum_n \Psi_n^\dag\psi_n$ of the Hamiltonian \eqref{eq:LatticeHopping} satisfies the following properties:
\begin{subequations}
	\label{eq:LatticeRecurBoundary}
\begin{align}
		\vb\cdot\vGamma \psi_{n-1} + \vb^0\cdot\vGamma \psi_n
				+ \vb^\ast\cdot\vGamma \psi_{n+1} & = E \psi_n	&& n \geq 1
		,	\label{eq:LatticeRecursion}
	\\	\psi_0 & = 0
		.	\label{eq:LatticeBoundary}
\end{align}
\end{subequations}
An edge state $\psi$ is one for which $\psi_n$ is exponentially decreasing as a function of $n$.
Due to the translational invariance in the bulk, we use the ansatz
\begin{align}
	\psi_n = \sum_\mu u_\mu \lambda_\mu^n
		\label{eq:LatticeWavefunction}
\end{align}
in our solutions.%
	~\cite{DHLeeSurfaceBandTM81}%
	~\footnote{A solution of the form $u_\mu n \lambda_\mu^{n-1}$ is also possible, when $\lambda_\mu$ is a double root of \eqref{eq:LatticeQuartic}.
	In such situation, we can tackle the problem as a limiting case of two roots approaching each other: $\lambda_\mu' \rightarrow \lambda_\mu$.}
An edge solution requires that $|\lambda_\mu| < 1$ for all $\mu$ and that $\sum_\mu u_\mu = 0$.

For each decaying mode ($\lambda, u$) we have:~%
	\footnote{%
		We cannot cross off $\lambda$ on both sides because $\lambda = 0$ is a valid solution when $Bu_\mu = 0$.
	}
\begin{align}
	\lambda \left[ \vbeta(\lambda) \cdot \vGamma - E \right] u = 0 ,
		\label{eq:LatticeElambda}
\end{align}
where we have defined
\begin{align}
	\vbeta(\lambda) \equiv \lambda^{-1}\vb + \vb^0 + \lambda\vb^\ast .
\end{align}
Notice that when $\lambda$ has unit modulus, we recover the Bloch equation for the bulk (propagating) modes of the system:
\begin{align}
	\vbeta \left( e^{i\kperp} \right) = \vh(\kperp) .
\end{align}
Hence we need to find an energy $E$ and a set of $\lambda$'s all within the unit circle, such that their corresponding null vector $u$ defined by \eqref{eq:LatticeElambda} sum to zero [Eq.~\eqref{eq:LatticeBoundary}].

The outline of the proof is as follows.
First, we derive the particle-hole relationship between left edge and right edge modes.
Second, we establish the algebraic relations between the $\lambda$'s and $E$ of an edge state [Eq.~\eqref{eq:LatticeEdgeCondbeta}].
Third, we define complex functions $L, \bar{L}$ which represents the ellipse traced out by $\vbeta(e^{ik})$.
Next, we show (assuming an edge state exists) that the energy of an edge state is given by the displacement of the ellipse $|\vbperp|$.
We then proceed to prove Thm.~1a, the condition which governs the existence of an edge state (\textit{i.e.} when all $|\lambda_\mu|<1$).
Finally, we compute the edge states projectors [Eq.~\eqref{eq:LatticeProjector}] and determine the sign of the edge state energies, which completes the proof for Thm.~1b.

\subsection{Relation between left-right boundaries}

\textbf{Lemma.}
\textit{%
For every left edge state with energy $E$, there is a corresponding right edge state with energy $-E$, and vice versa.%
}

The recursion relation \eqref{eq:LatticeRecursion} and boundary condition \eqref{eq:LatticeBoundary} describe a semi-infinite system with a ``left'' edge.
We can write a similar system for the ``right'' edge simply changing the condition in \eqref{eq:LatticeRecursion} to $n \leq -1$ xor swapping $\vb$ with $\vb^\ast$ in the equation.
The first transformation amounts to finding a set of $\lambda$'s outside the unit circle, \textit{i.e.} $|\lambda_\mu|>1$, such that the edge wavefunction $\psi_n$ decays with decreasing $n$.  The second transformation is equivalent to taking the complex conjugate of $\vb$, and the two transformation are the same owing to the fact that $\vbeta^\ast(\lambda) = \vbeta(1/\lambda^\ast)$.

When the gamma matrices are the Pauli matrices, there is a charge conjugation operator $C = -i\sigma^y\mathcal{K}$ which takes $\vb\cdot\bm\sigma \rightarrow -\vb^\ast\cdot\bm\sigma$, where $\mathcal{K}$ is the complex conjugation operator.  Since $C \vbeta(\lambda)\cdot\bm\sigma C^{-1} = -\vbeta^\ast(\lambda^\ast)\cdot\bm\sigma$, $C$ turns a left edge state with energy $E$ into a right edge state with energy $-E$ and vice versa.

When the Dirac matrices are larger then $2\times2$, such a $C$ operator still exists, as it is always possible to find an antiunitary operator which flips the sign of three of the gamma matrices.
This operator $C$ will depend on what $\vb$ and $\vb^0$ are, meaning that $C$ is a function of $\vkpara$, making it a non-local operator.  However, the conclusion remains the same.

For the remainder of the proof, we will only be focused on left edge states.

\subsection{Algebraic relation between $\lambda_a,\lambda_b$ and $E$}
\label{sec:LatticeTransferProof2}

For any $E$, there are (at most) four possible $\lambda$'s satisfying \eqref{eq:LatticeElambda}, evident from squaring $\lambda\vbeta(\lambda)\cdot\vGamma$ to get the quartic equation
\begin{align}
	\lambda\vbeta(\lambda) \cdot \lambda\vbeta(\lambda) - E^2\lambda^2 = 0 .
		\label{eq:LatticeQuartic}
\end{align}
Note that if $\lambda$ is a root to this equation, then so is $1/\lambda^\ast$.  Hence there can be at most two solutions of for $\lambda$ within the unit circle, which we call $\lambda_a$ and $\lambda_b$.

The edge wavefunction takes the form $\psi_n = u_a \lambda_a^n + u_b \lambda_b^n$ with $u_a = -u_b$ to satisfy the hard-edge boundary condition \eqref{eq:LatticeBoundary}, where the coefficient $u_a$ is a (right) null vector of the matrix $\lambda_a(\vbeta(\lambda_a)\cdot\vGamma-E)$ and similarly for $u_b$.
It follows that the matrices $\lambda_a(\vbeta(\lambda_a)-E)$ and $\lambda_b(\vbeta(\lambda_b)-E)$ must share a non-zero null vector, or equivalently, any linear combinations of the two matrices must be non-invertible.
In other words, an edge state at energy $E$ exists only if the following conditions%
~\footnote{Equations~\eqref{eq:LatticeEdgeCondDeta}, \eqref{eq:LatticeEdgeCondDetb} are redundant, but useful to be written out explicitly.}
are satisfied:
\begin{subequations}
\begin{align}
		&\Det{ \lambda_a(\vbeta(\lambda_a)\cdot\vGamma - E) } = 0
		,	\label{eq:LatticeEdgeCondDeta}
	\\	&\Det{ \lambda_b(\vbeta(\lambda_b)\cdot\vGamma - E) } = 0
		,	\label{eq:LatticeEdgeCondDetb}
	\\	&\Det{c_a\lambda_a (\vbeta(\lambda_a)\cdot\vGamma - E)
			+ c_b\lambda_b (\vbeta(\lambda_b)\cdot\vGamma - E)} = 0
		,	\label{eq:LatticeEdgeCondDetc}
\end{align}
\end{subequations}
for arbitrary $c_a,c_b$ and for $|\lambda_a|, |\lambda_b| < 1$.
The converse statement is also true, as \eqref{eq:LatticeEdgeCondDeta} and \eqref{eq:LatticeEdgeCondDetb} implies that the ranks of the matrices $\lambda_{a,b} (\vbeta(\lambda_{a,b})\cdot\vGamma - E)$ are at most half their dimension (a property of gamma matrices).
The last equation \eqref{eq:LatticeEdgeCondDetc} means that the two matrices must share a right null vector or a left null vector.
In the former case we have a left edge state at energy $E$, and in the latter case we have a right edge state at energy $E$, which by our lemma implies a left edge state at energy $-E$.

We may rewrite the equations in a more useful form:
\begin{subequations}
	\label{eq:LatticeEdgeCondbeta}
	\begin{align}
			\lambda_a^2 E^2 & = \lambda_a^2 \vbeta(\lambda_a) \cdot \vbeta(\lambda_a) ,
		\\	\lambda_b^2 E^2 & = \lambda_b^2 \vbeta(\lambda_b) \cdot \vbeta(\lambda_b) ,
		\\	\lambda_a\lambda_b E^2 & = \lambda_a\lambda_b \vbeta(\lambda_a) \cdot \vbeta(\lambda_b) .
	\end{align}
\end{subequations}

\subsection{Introducing functions $L,\bar{L}$}
Similar to $\vb^0 = \vbpara + \vbperp$ and $\vh = \vhpara + \vbperp$, we decompose $\vbeta$ into components parallel and perpendicular to the 2D plane (1D if the ellipse is degenerate) spanned by $\{\vb^r,\vb^i\}$:
$\vbeta(\lambda) = \vbeta_\parallel(\lambda) + \vbperp$
Keep in mind that while $\vbperp$ is a real vector,
$\vbeta_\parallel(\lambda) = \lambda^{-1}\vb + \vbpara + \lambda\vb^\ast$ is generally a complex vector, unless $\lambda = e^{ik}$ has unit modulus.

We want to find complex functions which trace out the same ellipse as $\vbeta_\parallel(e^{ik})$ in the complex plane.
We first choose two real orthogonal unit vectors $\vvhat_1, \vvhat_2$ as a coordinate basis%
of the 2D plane.
Let 
\begin{align}\begin{split}
		L(\lambda) & = \vbeta_\parallel(\lambda)\cdot\vvhat_1
					+ i\vbeta_\parallel(\lambda)\cdot\vvhat_2
		,	
	\\	\bar{L}(\lambda) & = \vbeta_\parallel(\lambda)\cdot\vvhat_1
					- i\vbeta_\parallel(\lambda)\cdot\vvhat_2
		.	\label{eq:LatticeDefLLbar}
\end{split}\end{align}
The loci $L(e^{ik})$ and $\bar{L}(e^{ik})$ both trace out the ellipse in the complex plane identical to $\vbeta_\parallel(e^{ik})$, but with different orientations.
In general, $L(\lambda)$ and $\bar{L}(\lambda)$ are not conjugate pairs unless $\lambda$ lies on the unit circle.
Expanding their definitions, we can see that both $\lambda L(\lambda)$ and $\lambda \bar{L}(\lambda)$ are quadratic polynomials in $\lambda$:
\begin{align}\begin{split}
		L(\lambda) & = q\lambda + w + p\lambda^{-1} ,
	\\	\bar{L}(\lambda) & = p^\ast\lambda + w^\ast + q^\ast\lambda^{-1} .
\end{split}\end{align}
where $p=\vb \cdot (\vvhat_1+i\vvhat_2)$, $w=\vbpara \cdot (\vvhat_1+i\vvhat_2)$ and $q=\vb^\ast \cdot (\vvhat_1+i\vvhat_2)$.%

It is straightforward to show from \eqref{eq:LatticeDefLLbar} that
$\vbeta_\parallel(\lambda_a) \cdot \vbeta_\parallel(\lambda_b) = \frac{1}{2}(L_a \bar{L}_b + \bar{L}_a L_b)$, where $L(\lambda_a)$ is abbreviated as $L_a$, \textit{etc}.
Equations~\eqref{eq:LatticeEdgeCondbeta} become
\begin{subequations}
	\label{eq:LatticeEdgeCondlambdaL}
	\begin{align}
			\lambda_a^2 (E^2 - \Delta^2) & = \lambda_a^2 L_a \bar{L}_a
				,	\label{eq:LatticeEdgeCondlambdaLa}
		\\	\lambda_b^2 (E^2 - \Delta^2) & = \lambda_b^2 L_b \bar{L}_b
				,	\label{eq:LatticeEdgeCondlambdaLb}
		\\	\lambda_a\lambda_b (E^2 - \Delta^2)
			& = \tfrac{\lambda_a\lambda_b}{2} (L_a \bar{L}_b + L_b \bar{L}_a)
				,	\label{eq:LatticeEdgeCondlambdaLc}
	\end{align}
\end{subequations}
where $\Delta \equiv |\vbperp|$.

\subsection{Edge state energy}
In this portion, we show that the existence of an edge state requires: $E = \pm\Delta$, $|\lambda_{a,b}| < 1$, and either $L_a=L_b=0$ or $\bar{L}_a=\bar{L}_b=0$.  The converse statement is trivially true by inspecting Eq.~\eqref{eq:LatticeEdgeCondlambdaL}.
Hence if an edge state of the semi-infinite chain exists, we show here that it must have energy $E = \pm|\vbperp|$. (First half of Thm.~1b.)

Assuming that neither $\lambda_a$ or $\lambda_b$ are zero, then
\begin{subequations}
	\label{eq:LatticeEdgeCondL}
	\begin{align}
			E^2 - \Delta^2 & = L_a \bar{L}_a
			,		\label{eq:LatticeEdgeCondLa}
		\\	E^2 - \Delta^2 & = L_b \bar{L}_b
			,		\label{eq:LatticeEdgeCondLb}
		\\	E^2 - \Delta^2 & = \tfrac{1}{2} (L_a \bar{L}_b + L_b \bar{L}_a) .
	\end{align}
\end{subequations}
which we can combine to get
\begin{align}
	( L_a - L_b )( \bar{L}_a - \bar{L}_b ) = 0
	.	\label{eq:LatticeEdgeCondL2}
\end{align}
Equations~\eqref{eq:LatticeEdgeCondL} and \eqref{eq:LatticeEdgeCondL2} are simply reformulations of the recursion relation \eqref{eq:LatticeRecursion} and boundary condition \eqref{eq:LatticeBoundary}.
We now proceed to show that $E = \pm\Delta$.

\emph{Proof by contradiction}
Suppose that $E^2 - \Delta^2 \neq 0$.
Then $L_a, L_b, \bar{L}_a, \bar{L}_b$ are all non-zero.
Eq.~\eqref{eq:LatticeEdgeCondL2} implies that $L_a = L_b$ or $\bar{L}_a = \bar{L}_b$.  Eq.~\eqref{eq:LatticeEdgeCondLa} and \eqref{eq:LatticeEdgeCondLb} together means that one equality implies the other, and hence $L_a = L_b$ and $\bar{L}_a = \bar{L}_b$ are both true.
Here we have two polynomials with roots $\lambda_a$ and $\lambda_b$:
\begin{align*}
		q\lambda^2 + (w-L_a)\lambda + p & = 0 ,
	\\	p^\ast\lambda^2 + (w^\ast-\bar{L}_a)\lambda + q^\ast & = 0 ,
\end{align*}
which means that $p/q = q^\ast/p^\ast = \lambda_a\lambda_b$.  However, since $|\lambda_a\lambda_b| < 1$, we have a contradiction as $|p/q|$ cannot be less than one and greater than one at the same time.
Hence we have shown, should an edge state exist, it must have energy $E = \pm\Delta$.
Now \eqref{eq:LatticeEdgeCondLa} tells us that either $L_a$ or $\bar{L}_a$ is zero which combined with \eqref{eq:LatticeEdgeCondL2} leads to the desired result.

If one of $\lambda_a,\lambda_b$ is zero, say $\lambda_b = 0$
(which happens when the ellipse is a circle),
then the expressions simplify as $\lambda_b L_b = p$ and $\lambda_b \bar{L}_b = q^\ast$.
From \eqref{eq:LatticeEdgeCondlambdaLb}, either $\lambda_bL_b$ or $\lambda_b\bar{L}_b$ is zero.
From \eqref{eq:LatticeEdgeCondlambdaLc} we have either $L_a=L_b=0$ or $\bar{L}_a=\bar{L}_b=0$,
and in either case, \eqref{eq:LatticeEdgeCondlambdaLa} implies that $E^2-\Delta^2=0$.

If both $\lambda_a$ and $\lambda_b$ are zero (which happens when the circle is centered on the origin), then $\lambda=0$ must be a double root to the polynomial \eqref{eq:LatticeQuartic}.
In this case, we have a flat band, and it is much easier to refer back to \eqref{eq:LatticeRecurBoundary} and solve the system directly.
One easily finds that the statement about edge states holds.

\subsection{Existence of edge states}
Finally, we use the fact that $|\lambda_a|,|\lambda_b| < 1$ to determine when an edge mode is present.
Recall that there are four zeroes (and two poles) to the equation $L(\lambda)\bar{L}(\lambda) = 0$, and that at most two of the roots have modulus less than one.
An edge mode exists if either $L(\lambda)$ or $\bar{L}(\lambda)$ has both roots $\lambda$ within the unit circle.  (No edge mode exists if each function $L,\bar{L}$ has one root within the unit circle.)

We can compute the number of zeroes (inside the unit circle) of the function $L$ by the contour integral $\frac{1}{2\pi i}\oint\! \frac{L'(z)}{L(z)} dz$ along the unit circle, which computes the number of zeroes minus number of poles within the unit circle.
As $L(\lambda)$ has one pole (at $\lambda=0$), $L(\lambda)$ has two zeroes if and only if the ellipse $L(e^{ik})$ wraps around the origin counterclockwise, leading to an edge state.
On the other hand, if $L$ wraps around the origin clockwise, then $\bar{L}(e^{ik})$ wraps around the origin counterclockwise, and there are two zeroes for $\bar{L}(\lambda)$ within the unit circle which also leads to an edge state.
In the case where the ellipse $\vbeta_\parallel$ does not wrap the origin, then neither $L$ or $\bar{L}$ has two roots within the unit circle, and an exponentially decaying solution to the semi-infinite system does not exist.

This completes the proof for Theorem~1a, which relates the presence of edge states to the properties of the ellipse $\vbeta(e^{ik}) = \vh(k)$.

\subsection{Sign of edge state energy}

In this section, we determine whether the left edge state energy is $+\Delta$ or $-\Delta$, where $\Delta \equiv |\vbperp| \geq 0$.
This will complete the final statement of Thm.~1b.

Define $\vvhat_\perp$ as the unit vector parallel to $\vbperp$ (assume $\Delta > 0$).
Recall that $\vvhat_1, \vvhat_2$ are unit vectors used in the definitions of $L, \bar{L}$, and so the three unit vectors are mutually orthogonal.
Define the corresponding gamma matrices $\Gamma^x = \vvhat_1\cdot\vGamma, \Gamma^y = \vvhat_2\cdot\vGamma, \Gamma^\perp = \vvhat_\perp\cdot\vGamma$, which pairwise anticommute.
As $\vbeta_\parallel\cdot\vvhat_1 = \frac{1}{2}(L + \bar{L})$ and $\vbeta_\parallel\cdot\vvhat_2 = \frac{1}{2i}(L - \bar{L})$,
\begin{multline}
	\lambda(\vbeta(\lambda) \cdot \vGamma - E)
		= \lambda(\Delta\Gamma^\perp - E)
		\\	+ \frac{\lambda L(\lambda)}{2} ( \Gamma^x - i\Gamma^y )
			+ \frac{\lambda \bar{L}(\lambda)}{2} ( \Gamma^x + i\Gamma^y ) .
\end{multline}
The first term annihilates the projector $\frac{1}{2}(1 + \frac{E}{\Delta}\Gamma^\perp)$
while the second and third terms annihilate $\frac{1}{2}(1 + i\Gamma^x\Gamma^y)$ and $\frac{1}{2}(1 + i\Gamma^x\Gamma^y)$ respectively.

By inspection, the projector $P = \sum uu^\dag$ on to the edge state depends on whether $L$ or $\bar{L}$ has two zeroes inside the unit circle, as well as the energy $E$:
\begin{align}
	P_\pm =
		\begin{cases}
			\tfrac{1}{4}(1 \pm \Gamma^\perp)(1 - i\Gamma^x\Gamma^y)	&	L_a=L_b=0	\\
			\tfrac{1}{4}(1 \pm \Gamma^\perp)(1 + i\Gamma^x\Gamma^y)	&	\bar{L}_a=\bar{L}_b=0
		\end{cases}
	\;,	\label{eq:LatticeProjector}
\end{align}
where $P_\pm$ projects on to the edge states with energy $E = \pm\Delta$.

When the irreducible representations of $\vGamma$ are $2\times2$, the product $i\Gamma^x\Gamma^y$ must equal either $\Gamma^\perp$ or $-\Gamma^\perp$, hence either $P_+$ or $P_-$ must be zero.
This implies that an edge state occurs only at $\Delta$ or $-\Delta$ but not both.
For example, when $\Gamma^x\Gamma^y = i\Gamma^\perp$ and $L = 0$, then $P_+ = \frac{1}{2}(1+\Gamma^\perp), P_- = 0$ and there is exactly one edge state at energy $\Delta$.
In general, we can determine $E$ via the orientation of the ellipse $\vh$:
\begin{align}
	E = \vb^0 \cdot \frac{ \vb^r\times\vb^i }{ |\vb^r\times\vb^i| } ,
\end{align}
where the cross product is defined from the commutation algebra $\vb^r\times\vb^i = -\frac{i}{4}\Tr\big\{ \bm\sigma\, [\vb^r\cdot\bm\sigma, \vb^i\cdot\bm\sigma] \big\}$.
This result gives us Theorem~1b.

When the  irreducible representations of $\vGamma$ are $2^m \times 2^m$ with $m \geq 2$, there are edge states at both $\Delta$ and $-\Delta$.  As $P_+ + P_- = \frac{1}{2}(1 \mp i\Gamma^x\Gamma^y)$, there are a total of $2^{m-1}$ left edge states.  Notice that the projectors $P_+$ and $P_-$ are related by the similarity transformation $P_+ = \Gamma^w P_- \Gamma^w$, where $\Gamma^w$ anticommutes with $\Gamma^{x,y,\perp}$, and so there must be $2^{m-2}$ edge states at each energy.

\subsection{Effective surface Hamiltonian}
\label{sec:SurfaceEffHam}

We can use the edge state projector to construct the effective surface Hamiltonian.
The operator
\begin{align}
	P_s = P_+ + P_- = \tfrac{1}{2}(1 - i\Gamma^x\Gamma^y)
	\label{eq:ProjectorEdgeStates}
\end{align}
projects on to the edge states (assuming $\vvhat_1$ and $\vvhat_2$ are chosen such that $L = 0$).
Notice that it only depends on $\vb\cdot\vGamma$, the nearest-neighbor coupling and not the on-site potential:
\begin{align}
	\tfrac{1}{2}(1 - i\Gamma^x\Gamma^y)
		& = \frac{1}{2} \left( 1 - i\frac{ [ \vb^r\cdot\vGamma, \vb^i\cdot\vGamma ] }
				{ 2|\vb^r \times \vb^i| } \right)
			\notag\\
		& = \frac{1}{2} \left( 1 + \frac{ [ \vb\cdot\vGamma, \vb^\ast\cdot\vGamma ] }
				{ \frac{4}{\pi} (\textrm{Area of ellipse}) } \right) .
\end{align}
The effective surface Hamiltonian is
\begin{align}
	H_s & = E P_+ - E P_- + E_\infty(1 - P_s)			\notag
		\\	& = P_s(\vbperp\cdot\vGamma) + E_\infty(1 - P_s) ,
	\label{eq:EffSurfaceHam}
\end{align}
where $E_\infty \rightarrow \infty$ such that the low energy theory describes the surface states.

\section{Proof for continuum Hamiltonians}
\label{sec:ProofContinuumTM}

The proof of Theorem~3 is very similar to the proof of Theorem 1 using transfer matrices, and so we present here a condensed version of the proof.

Consider a semi-infinite system with $x \geq 0$, and either $\psi(0) = 0$ (Dirichlet) or $\psi'(0) = 0$ (Neumann) boundary condition.  The momentum parallel to the surface is a good quantum number, and so we fix $\vp_\parallel$ to get an effective 1D problem.
We seek a solution of the form%
\footnote{
	A summand of the form $\kappa e^{i\kappa }$ is also permissible, provided $\kappa$ is a double root of the polynomial \eqref{eq:ContinuumQuartic}.}
$\psi(x) = \sum_\mu e^{i\kappa_\mu x} u_\mu$, with $\Im{\kappa_\mu} > 0$.  Each pair ($\kappa, u$) satisfies:
\begin{align}
	(C^0 + C^1 \kappa + C^2 \kappa^2 - E) u = (\vh(\kappa)\cdot\vGamma - E) u = 0 .
		\label{eq:ContinuumEkappa}
\end{align}
Squaring $\vh\cdot\vGamma$ gives us the quartic equation
\begin{align}
	\vh(\kappa) \cdot \vh(\kappa) - E^2 = 0
		\label{eq:ContinuumQuartic}
\end{align}
with real coefficients.  Hence if $\kappa$ is a root, then $\kappa^\ast$ is also a root.
Once again, we have at most two solutions for $\kappa$ in the upper half of the complex plane, and so the wavefunction must take the form $\psi(x) = u_a e^{i\kappa_a x} + u_b e^{i\kappa_b x}$.
With either Dirichlet ($u_a = -u_b$) or Neumann ($\kappa_a u_a = -\kappa_b u_b$) boundary condition, we have $u_a \propto u_b$ and so $\vh(\kappa_a)\cdot\vGamma-E$ and $ \vh(\kappa_b)\cdot\vGamma-E$ share a null vector.
By a similar argument to that in Sec.~\ref{sec:LatticeTransferProof2}, the existence of an edge state is equivalent to
\begin{align}
		&\Det{c_a (\vh(\kappa_a)\cdot\vGamma - E)
			+ c_b (\vh(\kappa_b)\cdot\vGamma - E)} = 0
		,	\label{eq:ContinuumEdgeCondDetc}
\end{align}
for all $c_a, c_b$ and $\Im\kappa_a, \Im\kappa_b > 0$.
Equivalently, we have
\begin{align}
	E^2 & = \vh(\kappa_a) \cdot \vh(\kappa_a)
		= \vh(\kappa_b) \cdot \vh(\kappa_b)
		= \vh(\kappa_a) \cdot \vh(\kappa_b)
		.	\label{eq:ContinuumEdgeCondbeta}
\end{align}

We introduce the functions $L(\kappa), \bar{L}(\kappa)$:
\begin{align}\begin{split}
		L(\kappa) & = \vhpara(\kappa)\cdot\vvhat_1
					+ i\vhpara(\kappa)\cdot\vvhat_2 ,
	\\	\bar{L}(\kappa) & = \vhpara(\kappa)\cdot\vvhat_1
					- i\vhpara(\kappa)\cdot\vvhat_2 ,
			\label{eq:ContinuumLatticeDefLLbar}
\end{split}\end{align}
where $\vvhat_1, \vvhat_2$ form an orthonormal coordinate basis in the plane spanned by $\vc^1$ and $\vc^2$.
$L(\kappa), \bar{L}(\kappa)$ are quadratic polynomials in $\kappa$:
\begin{align}\begin{split}
		L(\kappa) & = (c^0_x + ic^0_y) + (c^1_x + ic^1_y)\kappa + (c^2_x + ic^2_y)\kappa^2	,
	\\	\bar{L}(\kappa) & = (c^0_x - ic^0_y) + (c^1_x - ic^1_y)\kappa + (c^2_x - ic^2_y)\kappa^2	.
			\label{eq:ContinuumLLbarQuadratic}
\end{split}\end{align}
When $\kappa = p$ is real, $L(p)$ and $\bar{L}(p)$ trace out the parabola $\vhpara(p)$ in the complex plane with opposite orientations.
Using the relation $\vhpara(\kappa_a) \cdot \vhpara(\kappa_b) = \frac{1}{2}(L_a \bar{L}_b + \bar{L}_a L_b)$, where $\bar{L}(\kappa_a)$ is abbreviated as $\bar{L}_a$, \textit{etc.}, Eq.~\eqref{eq:ContinuumEdgeCondbeta} become
\begin{subequations}
	\label{eq:ContinuumEdgeCondL}
	\begin{align}
			E^2 - \Delta^2 & = L_a \bar{L}_a
					\label{eq:ContinuumEdgeCondLa} ,
		\\	E^2 - \Delta^2 & = L_b \bar{L}_b
					\label{eq:ContinuumEdgeCondLb} ,
		\\	E^2 - \Delta^2 & = \tfrac{1}{2} (L_a \bar{L}_b + L_b \bar{L}_a) ,
	\end{align}
\end{subequations}
where $\Delta \equiv |\vcperp|$.
The equations combine to get
\begin{align}
	( L_a - L_b )( \bar{L}_a - \bar{L}_b ) = 0 ,
		\label{eq:ContinuumEdgeCondL2}
\end{align}
Equations~\eqref{eq:ContinuumEdgeCondLa}, \eqref{eq:ContinuumEdgeCondLb} and \eqref{eq:ContinuumEdgeCondL2} together with $\Im\kappa_{a,b} > 0$ are true if and only if an edge state exists at energy $\pm E$.

We construct a proof by contradiction showing that $E = \pm\Delta^2$.  Suppose $E^2 \neq \Delta^2$, then all of $L_a, L_b, \bar{L}_a, \bar{L}_b$ are nonzero.  Equating \eqref{eq:ContinuumEdgeCondLa} and \eqref{eq:ContinuumEdgeCondLb} gives $L_a/L_b = \bar{L}_a/\bar{L}_b$, and combining with \eqref{eq:ContinuumEdgeCondL2} implies $L_a = L_b$ and $\bar{L}_a = \bar{L}_b$.
Hence the polynomials $L(\kappa)-L_a$ and $\bar{L}(\kappa)-\bar{L}_a$ have identical roots ($\kappa_a$ and $\kappa_b$).  The sum of the roots $\kappa_a + \kappa_b$ must lie in the upper half plane, and it is equal to $(c^1_x+ic^1_y)/(c^2_x+ic^2_y)$ and $(c^1_x-ic^1_y)/(c^2_x-ic^2_y)$ from \eqref{eq:ContinuumLLbarQuadratic}.  This leads to a contradiction as the expressions are complex conjugate pairs.
Therefore, an edge state requires $E = \pm\Delta$ and either $L_a=L_b=0$ or $\bar{L}_a=\bar{L}_b=0$.

Finally we impose the condition that $\Im\kappa_a, \Im\kappa_b > 0$.
There are no poles in the function $L(\kappa)$, and so the number of zeroes in the upper half plane is given by $\frac{1}{2\pi i}\oint\! \frac{L'(z)}{L(z)} dz$, integrated along the real line from $-R$ to $R$ and closed on the upper half plane $R e^{i\theta}$ for $0 \leq \theta \leq \pi$, where $R$ is taken to infinity.
Assuming that $\vc^2$ is non-zero and so $L$ is a quadratic function of $z$, the contour of $L(z)$ for $z = Re^{i\theta}$ always wraps the origin by $2\pi$ radians counterclockwise.

Hence $L(\kappa)$ has two roots in the upper half plane if and only if the parabola $L(p)$ wounds around the origin counterclockwise.  Similarly, $L(\kappa)$ has zero roots (so $\bar{L}(\kappa)$ has two roots) in the upper half plane if the parabola wounds around the origin clockwise.
An edge state exists in both these cases, which occur when the origin lies in the concave side of $\vhpara(p)$.
If the origin is not in the concave side of the parabola $\vhpara(p)$, then $L(\kappa)$ and $\bar{L}(\kappa)$ only have one root in the upper half plane and the system has no edge states.
This completes the proof for Theorem~3a.

To determine the sign of the edge states, we construct the projectors for $E = \pm\Delta$.  The projectors in the continuum case is identical to that of the lattice case \eqref{eq:LatticeProjector}, hence by the same argument used for Thm.~1b, we can prove Theorem~3b.
\begin{align}
	E = \vc^0 \cdot \frac{ \vc^1\times\vc^2 }{ |\vc^1\times\vc^2| } .
\end{align}
In addition, one may also derive the effective surface Hamiltonian:
\begin{align}
	H_s & = E P_+ - E P_- + E_\infty(1 - P_s)			\notag
		\\	& = P_s(\vcperp\cdot\vGamma) + E_\infty(1 - P_s) ,
	\label{eq:ContinuumEffSurfaceHam}
\end{align}

\bibliography{EdgeStates}

\end{document}